\documentclass[a4paper,11pt]{article}

\usepackage{amsmath}
\usepackage{graphicx}
\usepackage{amsfonts}
\usepackage{amssymb}

\newtheorem{theorem}{Theorem}[section]
\newtheorem{rema}{Remark}[section]

\newtheorem{defi}[rema]{Definition}
\newtheorem{lemma}[rema]{Lemma}

\newcommand{\sect}[1]{\setcounter{equation}{0}\section{#1}}

\newcommand{\bc}{\begin{center}}
\newcommand{\ec}{\end{center}}
\def\ba#1{\begin{array}{#1}\displaystyle}
\newcommand{\ea}{\end{array}}

\newcommand{\beq}{\begin{equation}}
\newcommand{\eeq}{\end{equation}}
\newcommand{\beqa}{\begin{eqnarray}}
\newcommand{\eeqa}{\end{eqnarray}}
\newcommand{\no}{\nonumber}
\newcommand{\n}{\nonumber\\}
\newcommand{\bi}{\begin{itemize}}
\newcommand{\ei}{\end{itemize}}

\def\lt#1{\left#1}
\def\rt#1{\right#1}
\def\t#1{\tilde{#1}}
\def\h#1{\hat{#1}}
\def\b#1{\bar{#1}}
\def\frc#1#2{\frac{#1}{#2}}

\newcommand{\p}{\partial}

\newcommand{\bra}{\langle}
\newcommand{\ket}{\rangle}
\newcommand{\Z}{{\mathbb{Z}}}

\newcommand{\R}{{\mathbb{R}}}
\newcommand{\C}{{\mathbb{C}}}
\newcommand{\hC}{{\hat{\mathbb{C}}}}

\newcommand{\uD}{{\mathbb{D}}}

\newcommand{\Or}{{\cal O}}

\newcommand{\ep}{\epsilon}

\newcommand{\id}{{\rm id}}

\newcommand{\halmos}{\rule{1ex}{1.4ex}}
\newcommand{\eproof}{\hspace*{\fill}\mbox{$\halmos$}}
\newcommand{\proof}{{\em Proof.\ }}

\newcommand{\dom}{{\Upsilon}}

\newcommand{\ev}{{\cal E}}

\newcommand{\tou}{{\cal X}}

\newcommand{\supp}{{\rm supp}}

\def\cl#1{\overline{#1}}

\newcommand{\tto}{\twoheadrightarrow}

\newcommand{\cT}{{\cal T}}
\newcommand{\tE}{{\tt E}}
\newcommand{\tX}{{\tt X}}
\newcommand{\tY}{{\tt Y}}
\newcommand{\tZ}{{\tt Z}}
\newcommand{\tT}{{\tt T}}
\newcommand{\tU}{{\tt U}}
\newcommand{\tI}{{\tt I}}
\newcommand{\Supp}{{\rm Supp}}

\newcommand{\sqa}{\square}

\begin{document}

\begin{titlepage}

\begin{center}
{\Large {\bf Hypotrochoids in conformal restriction systems and Virasoro descendants}

\vspace{1cm}

Benjamin Doyon}

Department of Mathematics, King's College London\\
Strand, London, U.K.\\
email: benjamin.doyon@kcl.ac.uk

\end{center}

\vspace{1cm}

\noindent 
A conformal restriction system is a commutative, associative, unital algebra equipped with a representation of the groupoid of univalent conformal maps on connected open sets of the Riemann sphere, along with a family of linear functionals on subalgebras, satisfying a set of properties including conformal invariance and a type of restriction. This embodies some expected properties of expectation values in conformal loop ensembles CLE$_{\kappa}$ (at least for $8/3< \kappa \leq 4$). In the context of conformal restriction systems, we study certain algebra elements associated with hypotrochoid simple curves (a family of curves including the ellipse). These have the CLE interpretation of being ``renormalized random variables'' that are nonzero only if there is at least one loop of hypotrochoid shape. Each curve has a center $w$, a scale $\ep$ and a rotation angle $\theta$, and we analyze the renormalized random variable as a function of $u=\ep e^{i\theta}$ and $w$. We find that it has an expansion in positive powers of $u$ and $\b u$, and that the coefficients of pure $u$ ($\b u$) powers are holomorphic in $w$ ($\b w$). We identify these coefficients (the ``hypotrochoid fields'') with certain Virasoro descendants of the identity field in conformal field theory, thereby showing that they form part of a vertex operator algebraic structure. This largely generalizes works by the author (in CLE), and the author with his collaborators V. Riva and J. Cardy (in SLE$_{8/3}$ and other restriction measures), where the case of the ellipse, at the order $u^2$, led to the stress-energy tensor of CFT. The derivation uses in an essential way the Virasoro vertex operator algebra structure of conformal derivatives established recently by the author. The results suggest in particular the exact evaluation of CLE expectations of products of hypotrochoid fields as well as non-trivial relations amongst them through the vertex operator algebra, and further shed light onto the relationship between CLE and CFT.
\vfill

{\ }\hfill 21 September 2012

\end{titlepage}

\tableofcontents

\sect{Introduction}

Two-dimensional conformal field theory (CFT) \cite{BPZ,Gins,DFMS97} has been extremely successful at describing critical phenomena in statistical models. The use of conformal invariance in the context of local quantum field theory, in particular thanks to the large algebraic structures that arise (like the Virasoro vertex operator algebra \cite{LL04}), has produced a wealth of exact results for critical exponents and universal amplitudes.

Recently, a new picture of CFT has emerged thanks to the advent of Schramm-Loewner evolution (SLE) \cite{S00}, and conformal loop ensembles (CLE) \cite{Sh06,ShW07}  (for reviews on SLE and related aspects, see for instance \cite{WSLE,LSLE,KNSLE,Cardy05,BBSLE,GSLE,RBGWSLE}). These provide measures for random non-intersecting and self-avoiding curves and loops in regions of the plane, with properties of conformal invariance. They are expected to describe the same universality classes of critical behaviors as those given by CFT. However, they produce a very different description than that of local QFT, putting emphasis on geometric objects (curves and loops) and measures rather than algebraic structures. It is very interesting to try to understand the relation between this geometric, measure description and the CFT's algebraic structures.

We introduce the notion of conformal restriction systems which captures expected properties of CLE expectation functionals. In this context (but here expressed in the CLE language), we study ``renormalized random variables'' supported on configurations where at least one loop takes certain shapes. Such a renormalized random variable is akin to a distribution on the space of configurations, which ``forces'' there to exist at least one loop with a given shape. We analyze the cases where the shapes are simple curves of hypotrochoid type. Hypotrochoids are particular curves formed by rolling a disk on the inside of a bigger circle, and taking the trace of a fixed point on the disk. When the circumference of the disk fits an integer number of times in the circumference of the circle, this is a family of simple curves containing the ellipse, described by
\[
	w+\ep e^{i\theta}(be^{i\alpha} + b^{-1} e^{-i\alpha}):\alpha\in[0,2\pi)
\]
for $k=2,3,4,\ldots$ and $b>(k-1)^{1/k}$. We look at the resulting renormalized random variables as functions of $u=\ep e^{i\theta}$ and $w$, where $\ep$ is the scale of the curve, $\theta$ its counter-clockwise rotation angle and $w$ its center. We find an asymptotic expansion in positive powers of $u$ and $\b u$, and we show that the coefficients of integer powers of $u$ and $\b u$ are holomorphic in $w$ and $\b w$, respectively, and independent of the parameter $b$, and that they can be identified with certain Virasoro descendants of the identity field in CFT.  These coefficients can be seen as ``ultra-local'' renormalized random variables, and are referred to as hypotrochoid fields.

This generalizes the works \cite{DRC06,DTCLE} where the case of the ellipse, at order $u^2$, was studied, and related to the stress-energy tensor. Our work shows that hypotrochoid fields fall into a vertex operator algebra structure: expectations of products of hypotrochoid fields can be evaluated using the Virasoro vertex operator algebra. Our results hold for all central charges between 0 and 1, and for expectations with any number of hypotrochoid fields, on any finitely connected region whose boundary components, if any, are Jordan curves. In particular, we generalize the results of \cite{DTCLE}, including those related to the relative partition function, to such finitely connected regions. Since our results hold in the context of conformal restriction systems, they may be applicable beyond the scope of CLE.

The paper is organized as follows. In Section \ref{sectContext} we review the main ideas underlying CFT and CLE, and their relation with statistical models, in particular $O(n)$ loop models, at criticality. We also introduce some notions useful to express our results, and fundamental in the context of conformal restriction systems. In Section \ref{sectResults}, we describe our main results. In Section, \ref{sectAnalysis} we analyze these results in the light of the Virasoro vertex operator algebra. In Section \ref{sectCRS}, we define the notion of conformal restriction systems, in which the results of this paper hold; we derive some of their properties; and we make connections with CLE and with results of \cite{DTCLE}. In Section \ref{sectProofs}, we provide the proofs. Concluding remarks are given in Section \ref{sectConclusion}.

\sect{Context} \label{sectContext}

\subsection{$O(n)$ loop models}

For definiteness, consider a hexagonal lattice on the plane, and denote by $C$ the set of positions (the ``sites'') of the centers of the faces. Consider a statistical model whose configurations are functions $f$ on $C$, valued in $\{-1,1\}$, constrained to take the value 1 on all sites that do not lie on some finite domain $A$. If the measure (or Boltzmann weight) is defined by giving a weight $x$ to every pairs of neighboring sites with opposite values, a weight 1 to all other pairs, and by multiplying all these, then the result is the well-known Ising statistical model on the hexagonal lattice in $A$ with fixed boundary conditions. At $x=1/\sqrt{3}$, the model is critical. This means in particular that, after taking the thermodynamic limit where $A$ is scaled up to the whole plane, correlations between the values of $f$ at different sites of the lattice decay algebraically as the Euclidean distance between the sites increases.

Let $E$ be the union of all edges of the hexagonal lattice, and $E_A\subset E$ the subset formed by every edge that borders at least one face whose center lies in $A$. There is a one-to-one map between the above space of configurations of $f$ on $C$ and the space of disjoint, self-avoiding loops on $E_A$. Indeed, one simply associate to every configuration the set of edges separating opposite signs. It is easy to see that every such set of edges yields disjoint self-avoiding loops. If $L$ is the total number of such edges, then the Ising-model Boltzmann weight is $x^L$. There is a one-parameter family of statistical models that naturally generalizes the Ising model. They are defined by the Boltzmann weight $x^L n^N$, where $N$ is the total number of loops, and $0<n\leq2$. These models are expected to be critical at the $n$-dependent values $x=1/\sqrt{2+\sqrt{2-n}}$ \cite{N82}. These are the so-called $O(n)$ loop models (see for instance \cite{Cardy05} for an overview of these and other related models). These are the models on which we will base our intuition below.

\subsection{Scaling limit and CFT}\label{ssectscalCFT}

Every critical statistical model, with appropriate translational and rotational invariance, is expected to be described in the scaling limit by a model of conformal field theory (CFT) \cite{BPZ,Gins,DFMS97}. One way of expressing this more precisely, for instance in the $O(n)$ models above, is as follows. For every finite set $s\subset C$ of sites, consider the linear space ${\cal F}_s$ of observables supported on $s$ (complex functions of $f|_s$, the configuration on $s$). Assume $0\in C$. A local family of observables is a one-parameter family $\sigma=(\sigma_{\lambda,0}\in{\cal F}_s:\lambda\in[1,\infty))$ for some $s \ni 0$. Let us construct their translates to other sites $(\sigma_{\lambda,w}:\lambda\in[1,\infty))$, $w\in C$, and let us denote $\sigma_{\lambda}(w) := \sigma_{\lambda,[w]}$ for $w\in \C$, where $[w]\in C$ is the nearest point in $C$ to $w$. Let us further denote by $\mathbb{E}(\cdot)_A$ the statistical expectation. Then there exists a set ${\cal G}_{\rm stat}$ of local families $\sigma$ such that every limit
\[
	\lim_{\lambda\to\infty} \mathbb{E}(\sigma_\lambda^{(1)}(\lambda w_1)
	\cdots \sigma_\lambda^{(n)}(\lambda w_n))_{\lambda A}
\]
exist for every $\sigma^{(1)},\ldots,\sigma^{(n)}\in {\cal G}_{\rm stat}$, and for every pairwise non-coincident $w_j\in A$. The result of all such limits is the scaling limit of the statistical model. An example in the Ising model is the family described by $\sigma_\lambda^{\rm spin}(w)=\lambda^{1/8} f([w])$, representing the scaling limit of the spin variable. The set ${\cal G}_{\rm stat}$ can naturally be taken to form a linear space.

The scaling limit is expected to have the structure of a local conformal field theory. That is, there is a CFT model such that the linear space ${\cal G}_{\rm stat}$ gives rise to the linear space ${\cal G}_{\rm CFT}$ of CFT local fields: for every family $\sigma\in {\cal G}_{\rm stat}$ there exists a CFT local field $\Or\in{\cal G}_{\rm CFT}$, $\sigma\equiv \Or$, such that
\[
	\lim_{\lambda\to\infty} \mathbb{E}(\sigma_\lambda^{(1)}(\lambda w_1)
	\cdots \sigma_\lambda^{(n)}(\lambda w_n))_{\lambda A}
	= \bra\Or^{(1)}(w_1)\cdots \Or^{(n)}(w_n)\ket_A
\]
for $\sigma^{(j)}\equiv \Or^{(j)}$, where $\bra\cdots\ket_A$ is the CFT correlation function on the domain $A$ (here with ``fixed'' conformal boundary condition).

The CFT correlation functions are universal: changing the actual measure on the lattice by ``irrelevant terms'' does not affect the result of the scaling limit. Taking a different regular lattice is also expected to preserve the scaling limit. Hence every CFT model represent the scaling limit of a whole universality class of statistical models.

Further, the structure of a local conformal field theory means that every field $\Or$ can be described in terms of vertex operator algebras and their modules (\cite{BPZ,LL04}). This description characterizes the universality class. Every vertex operator algebra contains the Virasoro vertex operator algebra, which includes the usual Virasoro algebra \cite{LL04}. A particularly important quantity is the central charge $c$ of the Virasoro algebra; this is one characteristic of the universality class. In the above connection between the $O(n)$ model and CFT, it is expected to be related to the quantity $n$ as
\beq\label{cn}
	c = \frc{(2-3y)(4y-1)}{1-y},\quad\cos(2\pi y) = -\frc{n}2,\quad \frc14< y\leq \frc12.
\eeq
As an example, in the case of the Ising model, $n=1$, with central charge $c=1/2$, the family $\sigma_\lambda^{\rm spin}(w)$ corresponds, conjecturally, to the spin field $\Or(w)$, which is a Virasoro irreducible-module highest-weight vector for the left-moving and right-moving copies of the Virasoro vertex operator algebra, of dimensions $(1/16,1/16)$ \cite{BPZ,Gins,DFMS97}.

\subsection{Scaling limit and CLE}\label{ssectscalCLE}

On the other hand, it is expected that the critical $O(n)$ statistical models, seen as models for random loops on $E_A$, converge, in a certain scaling limit, to conformal loop ensembles (CLE) \cite{Sh06,ShW07}. This convergence is in the sense of an appropriate topology on the space of measures \cite{Smi1,Smi2,Smi3}, and here the scaling limit is that where the mesh size is made smaller keeping the domain $A$ fixed, $E_A\mapsto (\lambda^{-1} E)_A$ with $\lambda\to\infty$. CLE form a one-parameter family of measures, with certain defining properties including conformal invariance. They are measures on the set of loop configurations (``loop soups''): every configuration is a set of countably-many non-intersecting and non-self-intersecting loops lying in $A$. They have mostly been studied on simply connected hyperbolic regions $A$ (i.e.~conformal to the disk), but here we will think of CLE on general, possibly multiply (but finitely) connected regions $A$ of the Riemann sphere $\hC = \C\cup \{\infty\}$, including $\hC$ itself. With respect to CLE measures, there is almost surely infinitely-many loops lying in $A$, but for any fixed $r>0$, there is almost surely a finite number of loops whose diameter (smallest covering disk) is larger than $r$. The CLE parameter is conventionally taken as $\kappa\in(8/3,8]$. This parameter can be seen as describing the almost-sure fractal dimension of the loops, $1+\kappa/8$. The $O(n)$ models described above are expected to converge to CLE in the dilute regime, $\kappa\in(8/3,4]$, where the loops are almost surely disjoint and self-avoiding (i.e.~simple). There, the relation between $n$ and $\kappa$ is expected to be
\beq\label{kappan}
	\kappa = \frc2{1-y}
\eeq
for $y$ as in (\ref{cn}). The convergence statement, in the dilute regime, has been proven in the case of the Ising model ($\kappa=3$) only \cite{Smi2,Smi3}. The dense regime, $\kappa\in(4,8]$ \cite{Sh06,ShW07}, is, in a sense, ``dual'' to the dilute regime; we will not discuss it here.

As $\kappa\to8/3$, $n\to 0$, and the ``density'' of loops in CLE decreases. At $\kappa=8/3$, there are no loops, hence the configuration set has only one element (the empty set). But it is possible to make the theory non-trivial, both in the scaling limit and in the $O(n)$ lattice model, by modifying the boundary conditions. Consider, in the set of sites not lying in $A$, those sites that have at least one neighbor lying in $A$. These are the boundary sites. If the boundary conditions are such that boundary sites are divided into two contiguous sets, one of which is fixed to $1$ and the other to $-1$, then there will be in every configuration one curve starting and ending on the boundary, where the two boundary sets join. In the scaling limit, the measure on this curve is expected to converge to the measure given by Schramm-Loewner evolution (SLE) \cite{S00} on $A$, with parameter $\kappa$ (again the proof exists in the Ising case). In particular, this makes sense also at $\kappa=8/3$, where it is the scaling limit of a self-avoiding random walk (SLE exists for all $\kappa\in [0,8]$). It is possible to generalize this to multiple curves, by having more than two jumps along the boundary sites. This gives rise to multiple SLE, or, considering the loops as well, to an ``augmented'' CLE, with additional random simple curves, disjoint from each other and from the loops, starting and ending on the boundary. We will only discuss the application of our results to the usual CLE with $\kappa\in(8/3,4]$, but we expect them to apply also in the augmented situation and $\kappa=8/3$.

\subsection{Relation between CLE and CFT}\label{ssectCLECFT}

CLE measures are believed to describe the same universality classes as do CFT models, with a relation between $c$ and $\kappa$ obtained by combining (\ref{cn}) and (\ref{kappan}),
\beq\label{ckapp}
	c = \frc{(6-\kappa)(3\kappa-8)}{2\kappa}.
\eeq
CLE and CFT have very different descriptions: measures on random loops in the former, local fields and vertex operator algebras in the latter. Hence it is natural to expect that there be certain ``non-local'' algebraic structures in CFT that relate to the random loops of CLE, and certain kinds of random variables in CLE that relate to the local fields of CFT. Here we are interested in the latter direction, in particular in attempting to recover (parts of) the vertex operator algebra structure in CLE.

CLE should contain much more information than do the well-studied rational models of CFT: it should be for instance possible to construct local fields in CLE that lie outside of the usual Kac table, in addition to all rational local fields. Further, since the CLE measure exists for all central charges between 0 and 1, CLE also relates to non-rational models of CFT, which have been much less studied.

The one structure of CFT that is present for all central charges is that of the Virasoro vertex operator algebra. Hence the associated local fields (the identity descendants, including the stress-energy tensor) should have natural constructions in CLE for every $\kappa$. One goal of this paper is to study such constructions. Two fundamental concepts stand out as fundamental in such constructions: locality and renormalization.

Locality in CFT, and more generally in quantum field theory (QFT), is a well-defined concept (see any textbook on QFT). In its strong version, the only one that we need here, it can be understood as meaning that a local field $\Or$ is the scaling limit of a family $\sigma$ of random variables supported on a finite number of sites, as in the discussion of Subsection \ref{ssectscalCFT}. Hence, a local field only ``feels'' an infinitesimal region (in the limit: a single point). In CLE, a related concept is that of {\em support} \cite{DTCLE} of a random variable. A support of a random variable $\tX$ is a closed subset $K$ of the Riemann sphere $\hC$ such that the value of $\tX$ is the same on all configurations whose set of loops intersecting $K$ is the same (here we see random variables in general as functions on the space of loop configurations in $\hC$). In general, any given random variable possesses a set of supports, as the definition does not provide a unique support. Most of our statements only require the existence of a support in particular regions (which we express by saying that the random variable ``is supported in'' that region). We will denote by $\Supp(\tX)$ the set of supports of $\tX$, and sometimes by $\supp(\tX)$ a particular support of $\tX$. A locality condition in CLE that is similar to, yet somewhat stronger than, that of CFT, which we will refer to as {\em ultra-locality}, would be that there be a support that is a single point (we say that the variable is supported on that point). However, it is easy to convince oneself that there are very few (if any) nontrivial random variables in CLE with single-point supports. In order to obtain ultra-local variables, and a structure of CFT in CLE, we need a concept of renormalization.

The definition of the local families $\sigma\in{\cal G}_{\rm stat}$ of observables in Subsection \ref{ssectscalCFT} consists in a particular way of performing a renormalization procedure in quantum field theory, leading to CFT. Indeed, the lattice model can be seen as a regularization scheme, and the scaling limit, including the definition of local families, is a renormalization procedure. Although, quite similarly, the lattice spacing is sent to zero in performing the scaling limit leading to CLE, no actual renormalization is performed: no local families are defined. Hence, it is natural to expect that in relating CLE random variables to CFT local fields, the random variables will need to be ``renormalized''. Such a renormalization in the CLE context was considered in \cite{DTCLE}. For our purposes, this goes as follows, paralleling the definition of local families of Subsection \ref{ssectscalCFT}. Consider a one-parameter family of random variable $\tX_\ep:\ep>0$, with $\ep$ playing the role of a regularization parameter. Consider also a linear space $\frak{X}$ of CLE random variables. Let us denote by $\mathbb{E}\big[\cdot\big]_A$ the expectation in CLE.
\begin{defi}\label{defiwl}
We say that the limit $\lim_{\ep\to0} \tX_\ep$ exists weakly locally with respect to $\frak{X}$ if the limit of expectation values $\lim_{\ep\to0}\mathbb{E}\big[\tX_\ep \tY\big]_A$ exists for every region $A$, and for every variable $\tY\in \frak{X}$ with the property that there are supports ${\supp}(\tX_\ep)$ and ${\supp}(\tY)$ such that $\lim_{\eta\to0} \cl{\cup_{\ep\in(0,\eta)} {\supp}(\tX_\ep)}$ and ${\supp}(\tY)$ are disjoint and lie in $A$.
\end{defi}
We identify the limit $\lim_{\ep\to0} \tX_\ep$ with the equivalence class of the family $\tX_\ep:\ep>0$ under the equivalence relation $(\tX_\ep) \sim (\tX_\ep')\Leftrightarrow \exists a>0,\,f,f' :(0,a)\to \R^+\,|\, \tX_{f(\ep)} = \tX_{f'(\ep)}' \;\forall \;\ep\in(0,a)$, where $f,f'$ are required to be right-continuous at $0$ and to satisfy $\lim_{\ep\to0} f(\ep) = \lim_{\ep\to0} f'(\ep) = 0$. All these notions generalize easily to the cases where $\ep$ is in other topological spaces. The limit $\tX:=\lim_{\ep\to0} \tX_\ep$ is a renormalized random variable. If $\supp(\tX_\ep)$ are supports for $\tX_\ep$, then by definition the set $\supp(\tX)=\lim_{\eta\to0} \cl{\cup_{\ep\in(0,\eta)} \supp(\tX_\ep)}$ is a support for $\tX$. With the weak-local convergence, expectation values provide us with the {\em weak-local topology}, and we may complete the space $\frak{X}$ with respect to it, in particular including sequences of renormalized random variables that converge weakly locally with respect $\frak{X}$ (it is easy to verify that the definition still makes sense). The completion will be denoted $\cl{\frak{X}}$. Renormalized random variables lie in this completion. In the following sections, we will also use other properties (holomorphicity, etc.) in the weak-local sense, always with the interpretation of the validity in expectation values with $\tY$ supported away.

We may now ask for ultra-local renormalized random variables: those that possess single-point supports. We expect that all Virasoro identity descendants in CFT, as well as, more generally, all fields associated to other eventual internal symmetries, be constructible in CLE (and generalizations) via ultra-local renormalized random variables; but in general, for other CFT fields corresponding to nontrivial modules, that this not be the case. Because of the possible relation with CFT, we will refer to ultra-local renormalized random variables as {\em fields}.

Given sequences $\tX_\ep, \tX_\ep',\ldots$ that converge weakly locally as $\ep\to0$, we may ask about the convergence of products $\tX_\ep \tX_\ep'\cdots$.
\begin{defi}\label{defics}
We say that a set ${\cal G}$ of renormalized random variables is a {\em consistent set} with respect to $\frak{X}$ if the multiple limit
\[
	\lim_{\ep\to0}\lim_{\ep'\to0}\cdots\;\mathbb{E}\Big[\tX_\ep
	\tX'_{\ep'}\cdots\tY\Big]_A
\]
exists and doesn't depend on the order of the individual limits, for every region $A$, for every $\{\tX,\tX',\ldots\}\subset{\cal G}$ and for every random variable $\tY\in\frak{X}$ such there are supports of $\tX,\tX',\ldots,\tY$ that are pairwise disjoint and lie in $A$.
\end{defi}
The result of the limit indeed does not depend on the members of the equivalence classes taken, and is denoted $\mathbb{E}\Big[\tX\tX'\cdots\tY\Big]_A$. Note that this is somewhat in parallel to the definition of local families ${\cal G}_{\rm stat}$. The set ${\cal G}$ can naturally be taken to form a linear space. Clearly we can form the space $\frak{X}_{\cal G}$ by linearly adding to $\frak{X}$ all products of elements of ${\cal G}$ with disjoint supports. Both Definitions \ref{defiwl} and \ref{defics} make sense with respect to $\frak{X}_{\cal G}$, and in particular every $\tX\in{\cal G}$ is a renormalized random variable (i.e.~the corresponding family is convergent weakly locally) with respect to $\frak X_{\cal G}$. The space $\frak{X}_{\cal G}$ is a quasi-algebra: there is a multiplication between elements that have disjoint supports.

A precise relation, which we would like to establish, between ultra-local CLE random variables and CFT fields is as follows. For a (possibly renormalized) random variable $\tY$ let us associate to it the (formal) CFT field $\Or_\tY$, its ``one-point functions'' being defined by $\mathbb{E}\big[\tY]_A = \bra\Or_\tY\ket_A$ for every region $A$ where the random variable is supported. Further, for a renormalized random variable $\tX$ supported on the point 0, we denote by $\tX(w)$ its translate to the point $w$.
\begin{defi}
A linear space ${\cal G}_{\rm CLE}$ of ultra-local renormalized random variables supported on the point 0 is a reconstruction of a linear space ${\cal G}_{\rm CFT}$ of CFT local fields with respect to the algebra or quasi-algebra $\frak{X}$, with a bijection ${\tt X}\equiv \Or: \tX\in{\cal G}_{\rm CLE},\;\Or\in{\cal G}_{\rm CFT}$, if (i) $\{\tX_1(w_1),\ldots,\tX_N(w_N)\}$ forms with respect to $\frak{X}$ a consistent set for every $\{\tX_1,\ldots,\tX_N\}\subset {\cal G}_{\rm CLE}$ and for every non-coincident $w_j$s, and (ii)
\beq\label{iden}
	\mathbb{E}\big[{\tt X}_1(w_1)\cdots {\tt X}_N(w_N) \,\tY\big]_{A} =
	\big\bra \Or_1(w_1)\cdots\Or_N(w_N)\,\Or_\tY\big\ket_A,\quad
	{\tt X}_j\equiv \Or_j\;\forall\;j
\eeq
for every $\{\tX_1,\ldots,\tX_N\}\subset {\cal G}_{\rm CLE}$, for every region $A$, for every non-coincident $w_j$s lying in $A$, and for every $\tY\in\frak{X}$ supported in $A\setminus \{w_1,\ldots,w_N\}$.
\end{defi}
Note that this implies that (\ref{iden}) holds for every $\tY\in\frak{X}_{{\cal G}_{\rm CLE}}$ supported in $A\setminus \{w_1,\ldots,w_N\}$.

Of course, only one-point functions $\bra\Or_\tY\ket_A$ of the formal field $\Or_\tY$ were defined. However, if ${\cal G}_{\rm CFT}$ is a linear space of local fields associated to symmetries, then there is a fundamental definition of the correlation functions involved in (\ref{iden}), from the knowledge of the symmetry relations for $\bra\Or_\tY\ket_A$. Hence at least in these cases, (\ref{iden}) is not an empty statement. The construction presented in this paper is associated to the conformal symmetry of CFT; hence it is the appearance of a Virasoro vertex operator algebra structure that allows us to identify the result of the left-hand side of (\ref{iden}) with CFT correlation functions as in the right-hand side.

In Section \ref{sectResults} the results we express are restricted to finitely connected regions $A$ whose boundary components, if any, are Jordan curves; it is clear that the concepts above can be restricted to these cases. Also, as mentioned, our results hold in the context of conformal restriction systems, as defined in Section \ref{sectCRS}, and we expect CLE expectation functionals to give rise to such a system.

\subsection{Relevant previous results}\label{ssectrelevant}

In a recent  work \cite{DTCLE}, the author found the CLE renormalized random variable corresponding to the (bulk) stress-energy tensor ${\cal T}(w)$ in CFT on simply connected domains. This was based on, and largely generalized, works \cite{FW03,DRC06} where similar results, for bulk and boundary fields, were obtained in SLE at $\kappa=8/3$ and other conformal restriction measures (but in more general domains).

Consider the ellipse centered at $w$, with eccentricity $e = 2b/(1+b^2)\in(0,1)$ (with $b>1$) and major semi-axis of length $\ep(b+1/b)>0$ at angle $\theta$ with respect to the positive real axis\footnote{Note the different normalization of $\ep$, and the shift in $\theta$ due to the factor $i$, as compared to the choice made in \cite{DRC06,DTCLE}.}:
\beq\label{ellipse}
	\p E_2(w,\ep,\theta,b):= \lt\{w+\ep e^{i\theta}(be^{i\alpha} + b^{-1} e^{-i\alpha}):\alpha\in[0,2\pi)\rt\}.
\eeq
Let $E_2(w,\ep,\theta,b)$ be the domain bounded by this ellipse and containing the point $w$. Let ${\tt E}_2(w,\ep,\theta,b)$ be a random variable that, in essence, takes the value one if the configuration is separated between the inside and the outside of the ellipse, and zero otherwise. This requires a renormalization in CLE, because this exact definition gives a random variable that is almost surely zero, due to the almost sure presence of infinitely many loops surrounding almost every point \cite{ShW07}. Let us then define ${\tt E}_2(w,\ep,\theta,b)$ more precisely as follows. Let $A$ be the annular domain $E_2(w,\ep,\theta;b)\setminus \cl{E_2(w,\ep,\theta,(1-\eta)b)}$, for some $\eta>0$ small enough. Note that  indeed $\cl{E_2(w,\ep,\theta,(1-\eta)b)}\subset E_2(w,\ep,\theta,b)$. Let ${\tt E}_2^\eta(w,\ep,\theta,b)$ be the random variable that takes the value one if there is at least one CLE loop that lies on $A$ and that separates the two components of $\hC\setminus A$, and zero otherwise. Then we define
\[
	{\tt E}_2(w,\ep,\theta,b) := \lim_{\eta\to0}
	\frc{{\tt E}_2^\eta(w,\ep,\theta,b)}{{\mathbb{E}\big[
	{\tt E}_2^\eta(0,1,0,b)}\big]_{\hC}}.
\]
According to \cite[Eq 4.9]{DTCLE}, under certain hypotheses, the limit exists weakly locally for CLE in any simply connected region. The support of $\tE_2(w,\ep,\theta,b)$ can be chosen to be the ellipse (\ref{ellipse}). This is a first renormalization step.

In \cite{DTCLE}, the reconstruction of the CFT stress-energy tensor was found to be
\beq\label{resell}
	\lim_{\ep\to0} \frc1{2\pi\ep^2}\int d\theta\,e^{-2i\theta}\, {\tt E}_2(w,\ep,\theta,b)
	\equiv {\cal T}(w).
\eeq
The left-hand side can be interpreted as a ``rotating spin-2 ellipse", and involves a second renormalization step. The resulting renormalized random variable on the left-hand side of (\ref{resell}) is supported on $w$. Note that this result holds for any $b$: the resulting local field is independent of $b$. The reconstruction was obtained by proving, up to certain hypotheses about the CLE measure, the conformal Ward identities, the boundary conditions, and the stress-energy tensor conformal transformation properties with some central charge $c'$, and by identifying $c'=c$ via a more intuitive CFT-related argument. The reconstruction was actually proven only in simply connected regions $A$, and only for single insertions of the stress-energy tensor.

The same reconstruction was found in SLE at $\kappa=8/3$ \cite{DRC06}, with rather ${\tt E}_2(w,\ep,\theta,b)$ the random variable that takes the value one if the SLE curve does not intersect the ellipse $\p E_2(w,\ep,\theta;b)$, and zero otherwise (this reconstruction is valid in any region $A$ and with any number of insertions). Note that in this case, no renormalization was necessary in the first step of the definition; this is understood as being a consequence of the fact that the central charge is 0 at $\kappa=8/3$ (see the discussion in \cite{DTCLE}). In the SLE case, we may take $b=1$, where the ellipse collapses to a segment of length $4\ep$ (in CLE this is not possible, because the definition involves the ellipse with $(1-\eta)b$).

\sect{Results}\label{sectResults}

The results presented here are proven in the context of {\em conformal restriction systems}, which we define fully in Section \ref{sectCRS}. A conformal restriction system is a commutative, associative algebra $\frak{X}$ with a notion of support for elements in $\frak{X}$ (a supported algebra), and linear functionals $\mathbb{E}\big[\cdot\big]_A$, for every region $A\subset\hC$ of the Riemann sphere, on the subalgebras of elements supported in $A$, with properties of conformal invariance and restriction. Below we will implicitly restrict ourselves to regions $A$ that are finitely connected and whose boundary components, if any, are Jordan curves. We expect that an appropriate choice of an algebra $\frak{X}$ of random variables with the notion of support described in the precious section, along with CLE expectation values, constitute a conformal restriction system. In a way, this puts on a more solid basis and generalizes the hypotheses made in \cite{DTCLE} about CLE, and allows us to provide precise statements which may be applied more generally. We will show that the hypotheses made about CLE in \cite{DTCLE} follow from the axioms of conformal restriction systems, hence that all results of \cite{DTCLE} are actually results in conformal restriction systems. The proofs for the theorems in this section will be provided in Section \ref{sectProofs}.

Below, we assume the we are given both the algebra $\frak{X}$ and the linear functionals $\mathbb{E}\big[\cdot\big]_A$, for every region $A$ as stated above, of a conformal restriction system, and we refer to the elements of $\frak{X}$ as random variables (because of the connection with CLE), from which renormalized random variables can be obtained as explained in the previous section.

As preliminaries, let $N$ be the closure of an annular region with Jordan boundaries. For every such $N$, let $\tI(N)$ be, in CLE, the random variable that takes the value 1 if there is at least one CLE loop lying in $N$ and separating the components of $\hC\setminus N$, and zero otherwise. See Section \ref{sectCRS} for a definition of $\tI(N)$ in the general context of conformal restriction systems (Point 1 of the axioms). We assume the following two facts (Points 2 and 3 of the axioms): for every Jordan curve $\alpha$, the limit
\beq\label{limalpha}
	\lim_{N\to \alpha} \frc{\tI(N)}{\mathbb{E}\big[\tI(N)\big]_\hC}=:\tE(\alpha)
\eeq
exists weakly locally with respect $\frak{X}$ (in an appropriate topology); and the space
\[
	{\cal G} = {\rm span}\{\tE(\alpha):\alpha\mbox{ Jordan curve}\}
\]
is a consistent set with respect to $\frak{X}$. These are natural generalizations of some of the hypotheses made in \cite{DTCLE}. This implies that we can extend $\frak{X}$ to $\frak{X}_{\cal G}$; since this extension is rather fundamental, we will denote it by
\[
	\h{\frak{X}}:=\frak{X}_{\cal G}.
\]

\subsection{Hypotrochoids}

The results reviewed in Subsection \ref{ssectrelevant} can be understood as follows. We consider a renormalized random variable $\tE_2(w,\ep,\theta,b)$ that restricts there being at least one CLE loop taking the shape of an ellipse, and we look at the second Fourier mode (in the angle of the major semi-axis with respect to the positive real axis) of this variable. The result is that in the (further renormalized) limit where the extent vanishes, we obtain the stress-energy tensor.

We may ask more general questions: what about the other Fourier modes, or the analytic structure of the variable $\tE_2(w,\ep,\theta,b)$ itself, say as a function of $w$ and $u=\ep e^{i\theta}$? Or more generally, what is the analytic structure of a variable that restricts there being one loop taking a given shape?

\begin{figure}
\bc
a.\includegraphics[width=3cm,height=2.07cm]{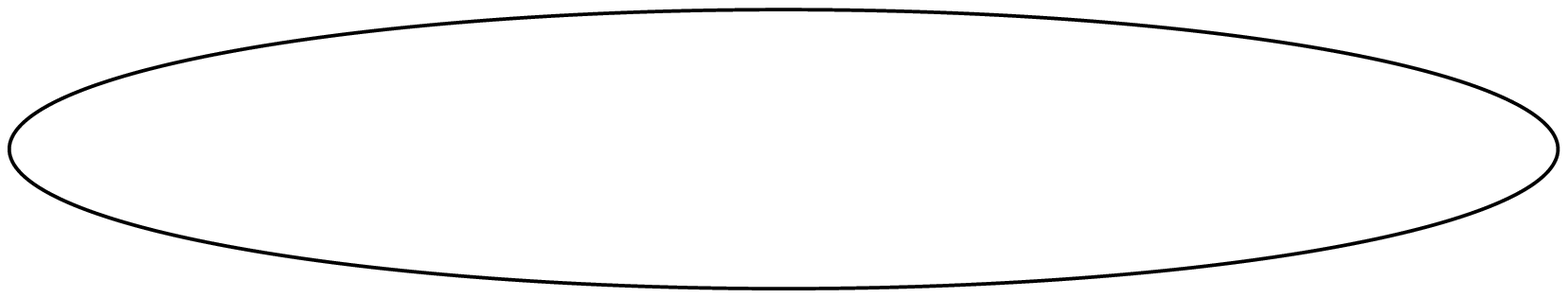}
b.\includegraphics[width=3cm,height=2.07cm]{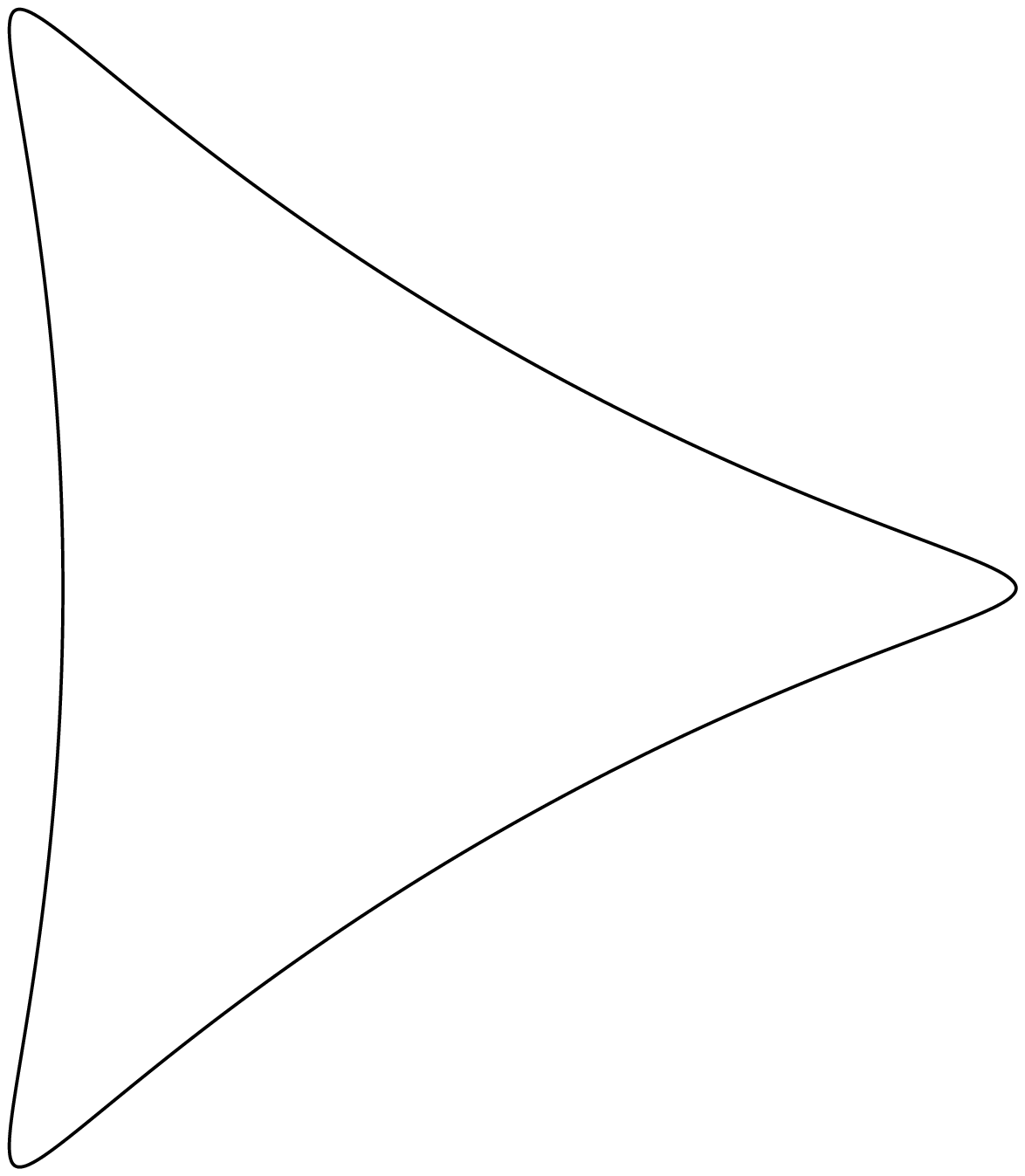}
c.\includegraphics[width=3cm,height=2.07cm]{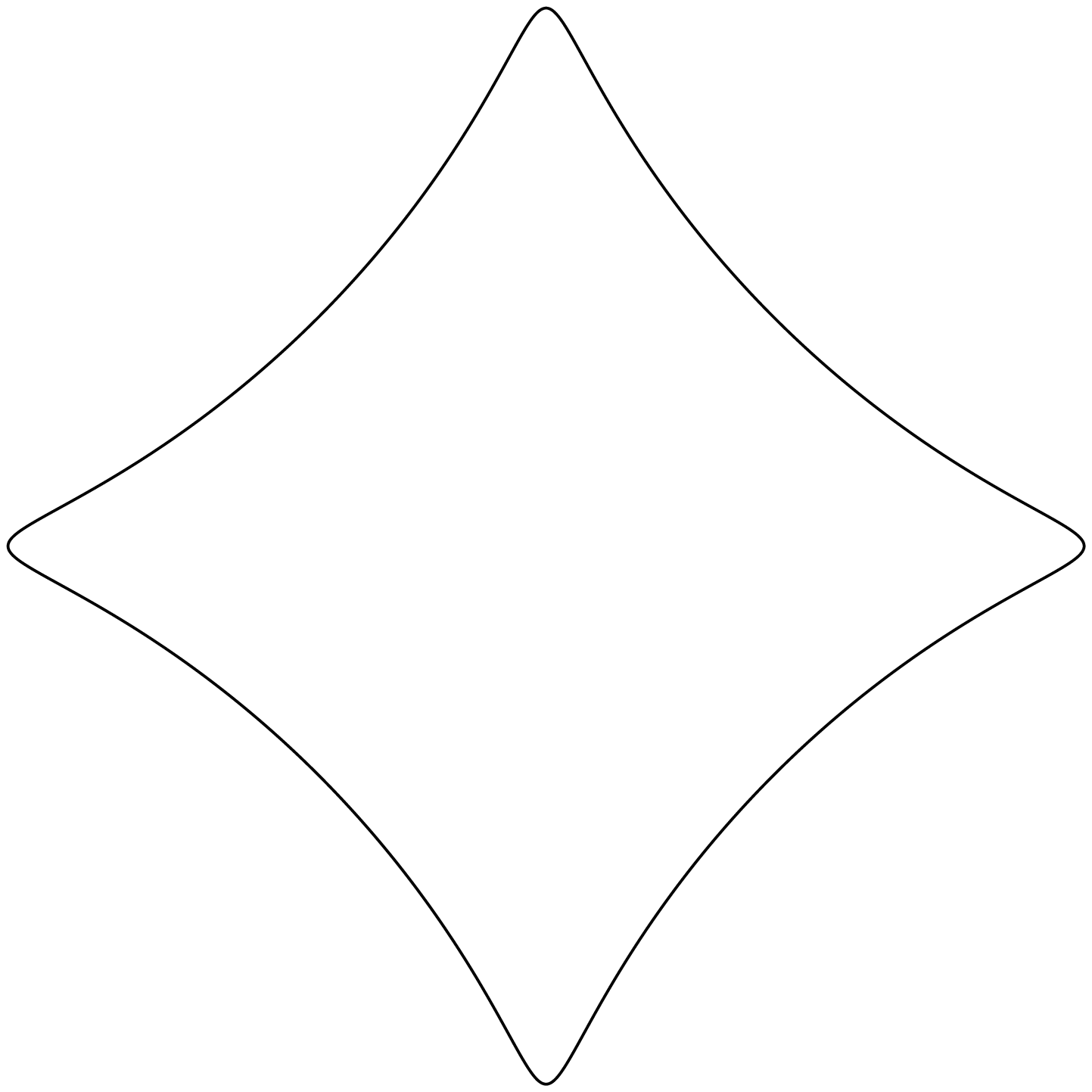}
d.\includegraphics[width=3cm,height=2.07cm]{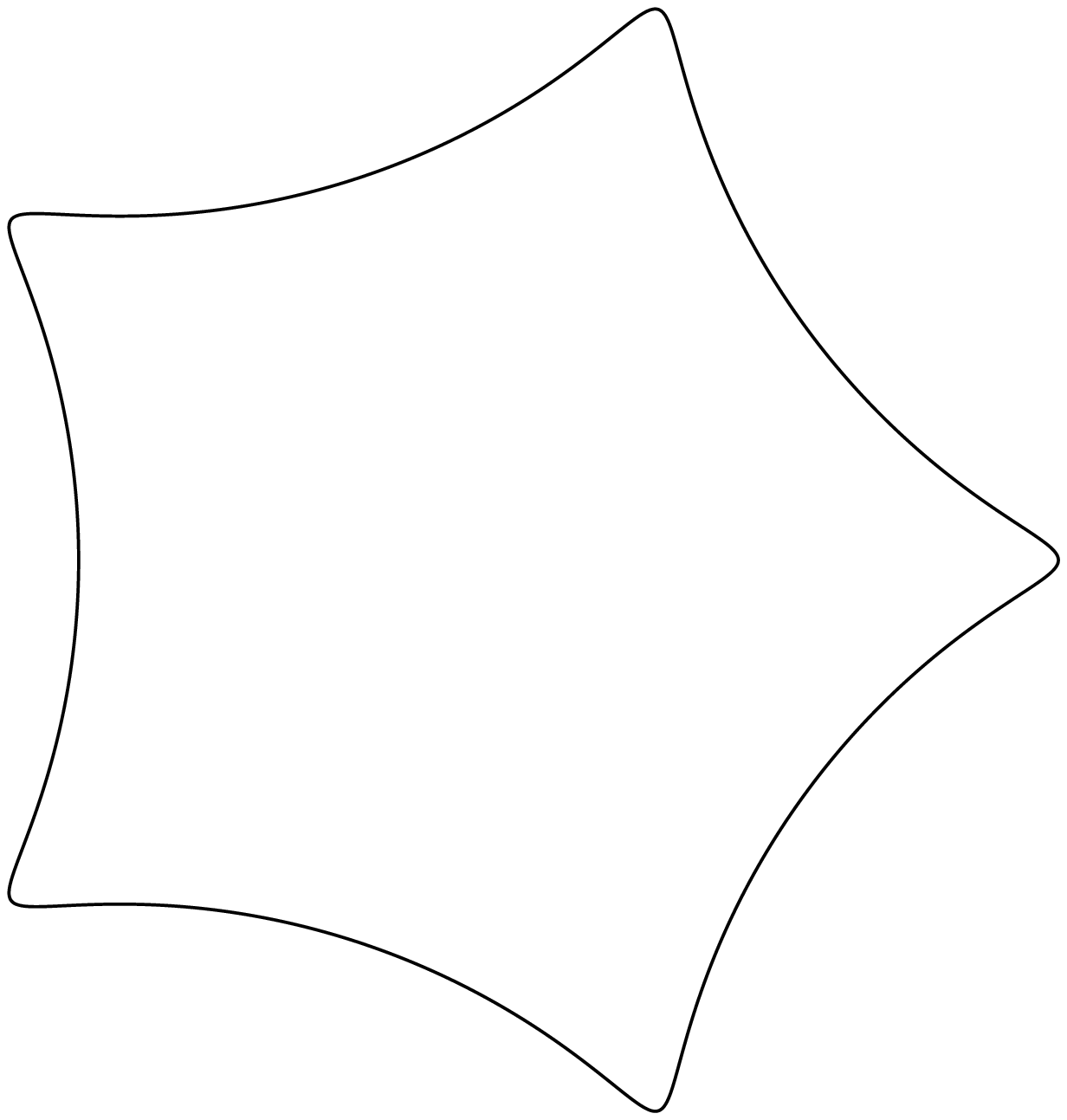}
\ec
\caption{The hypotrochoids corresponding to (a) $E_2$, (b) $E_3$, (c) $E_4$ and (d) $E_5$, all at angle $\theta=0$ (the horizontal is the real axis and the vertical the imaginary axis), and for generic (small) values of $b$.
}
\label{figcurves}
\end{figure}

The general question is rather complicated. We will restrict our attention to the following particularly interesting and symmetric family of shapes (simple closed curves in $\C$):
\beq\label{higher}
	\p E_k(w,\ep,\theta,b) := \lt\{w+\ep e^{i\theta}(be^{i\alpha} + b^{1-k} e^{(1-k)i\alpha}):\alpha\in[0,2\pi)\rt\}
\eeq
for $k=2,3,4,\ldots$, $w\in \C$, $\theta\in[0,2\pi)$ and $b>(k-1)^{1/k}$. At $k=2$, this is the ellipse (\ref{ellipse}), and in general for $k>2$, it is an hypotrochoid. It is a simple (for $b>(k-1)^{1/k}$) smooth curve centered at $w$ with dihedral $D_k$ symmetry about $w$ (see Figure \ref{figcurves}). The parameter $\ep$ is proportional to the diameter of the curve, and $\theta\;{\rm mod}\;2\pi/k$ is the positive angle between the positive real direction from $w$ and the first ``arm'' anti-clockwise. The arms become spikes (with $0$-angle corners) at $b = (k-1)^{1/k}$, where the curves (\ref{higher}) become hypocycloids. We define $E_k(w,\ep,\theta,b)$ as the domain bounded by (\ref{higher}) and containing $w$.

We will use the notation
\beq\label{Ek}
	\tE_k(w,\ep,\theta,b):=\tE(\p E_k(w,\ep,\theta,b)).
\eeq
Let $N_\eta(w,\ep,\theta,b):=\cl E_k(w,\ep,\theta;b)\setminus E_k(w,\ep,\theta;(1-\eta)b)$, for some $\eta>0$ small enough. Then it is immediate that
\beq\label{res1}
	{\tt E}_k(w,\ep,\theta,b) = \lim_{\eta\to0}
	\frc{\tI(N_\eta(w,\ep,\theta,b))}{\mathbb{E}\big[
	\tI(N_\eta(0,1,0,b))\big]_{\hC}}.
\eeq
This holds thanks to conformal invariance, $\mathbb{E}\big[\tE(N_\eta(0,1,0,b))\big]_{\hC} = \mathbb{E}\big[\tE(N_\eta(w,\ep,\theta,b))\big]_{\hC}$. That is, generalizing the results recalled above, we may define renormalized random variables $\tE_k(w,\ep,\theta,b)$ associated to hypotrochoids via a multiplicative renormalization that is independent of the geometric parameters $w,\ep,\theta$. The set of supports of the renormalized random variable $\tE_k(w,\ep,\theta,b)$ of course contains the closed curve (\ref{higher}).

In SLE$_{8/3}$, a corresponding definition would be to take ${\tt E}_k(w,\ep,\theta,b)$ as the random variable that takes the value one if the SLE curve does not intersect the curve $\p E_k(w,\ep,\theta;b)$, and zero otherwise.

\subsection{Asymptotic expansion}

As a function of $\theta$, the variable $\tE_k(w,\ep,\theta,b)$ should have the following Fourier expansion:
\[
	\tE_k(w,\ep,\theta,b) = \sum_{m\in\Z} e^{kmi\theta} {\tt e}_{k,m}(w,\ep,b).
\]
Further, ${\tt e}_{k,m}(w,\ep,b)$, for $m\neq 0$, should vanish as $\ep\to0$ (weakly locally). However, there is little else we can say {\em a priori}. 

Instead of using $\theta$ and $\ep$, let us consider the expansion of $\tE_k(w,\ep,\theta,b)$ in powers of the complex variables $u=\ep e^{i\theta}$ and $\b u = \ep e^{-i\theta}$. Then we have:
\begin{theorem}\label{res2}
With respect to $\h{\frak{X}}$, the renormalized random variable $\tE_k(w,\ep,\theta,b)$ has an asymptotic expansion in $u$ and $\b u$ of the following form, weakly locally, for every $w,b$:
\beq\label{asexp}
	\tE_k(w,\ep,\theta,b) \sim {\bf 1} + o((u\b u)^0)+ \sum_{m=1}^\infty \frc{u^{km}}{m!} \lt(\tT_{k,m}(w) + o((u\b u)^0)\rt) + \sum_{m=1}^\infty \frc{\b{u}^{km}}{m!} \lt(\b{\tT}_{k,m}(\b{w}) + o((u\b u)^0)\rt).
\eeq
The coefficients $\tT_{k,m}(w)$ and $\b \tT_{k,m}(\b{w})$ are independent of $b$; the former are holomorphic in $w$, the latter anti-holomorphic (weakly locally).
\end{theorem}
In particular, the expansion only contains {\em positive} powers of both $u$ and $\b u$. Further, the renormalized random variables $\tT_{k,m}(w)$ and $\b\tT_{k,m}(\b w)$ are supported on $w$. This is clear from the inverse relations, for instance
\beq
	\tT_{k,m}(w) = \lim_{\ep\to 0} \frc{m!}{2\pi\ep^{km}} \int d\theta \,e^{-kmi\theta} \,{\tt E}_k(w,\ep,\theta,b).
\eeq
We will refer to these as {\em hypotrochoid (anti-)holomorphic fields} or simply hypotrochoid fields.

\subsection{Virasoro descendants}

The stress-energy tensor is a Virasoro descendant of the identity field ${\bf 1}(w)$, as it can be expressed in the form ${\cal T}(w) = L_{-2}{\bf 1}(w)$. In the vertex operator language, this is $Y(L_{-2}{\bf 1},w)$, where $Y$ is the vertex-operator map and ${\bf 1}$ is the identity vector, highest weight of the identity module of the Virasoro algebra \cite{LL04}. The results reviewed in Subsection \ref{ssectrelevant} indicate that $\tT_{2,1}$ is a reconstruction of the stress-energy tensor, $\tT_{2,1}\equiv {\cal T}$ (we use $\tT_{k,m}:=\tT_{k,m}(0)$). We find a wide generalization of this: the space
\beq\label{GCLE}
	{\cal G}_{\rm CLE}:= {\rm span}\{\tT_{k,m}:k\geq 2,\;m\geq 1\}
\eeq
is a reconstruction of a certain chiral subspace ${\cal G}_{\rm CFT}$ of the identity sector of CFT. Similarly, the space $\b{\cal G}_{\rm CLE}:= {\rm span}\{\b\tT_{k,m}:k\geq 2,\;m\geq 1\}$ is a reconstruction of the (chiral) conjugate CFT subspace $\b{\cal G}_{\rm CFT}$.

The space ${\cal G}_{\rm CFT}$ is obtained from the following identity descendants in CFT:
\beq
	{\cal T}_{k,m} := \sum_{\lambda\in\Phi(m)} C_\lambda (k-1)^{m-|\lambda|} L_{-k\lambda_{|\lambda|}} \cdots L_{-k\lambda_2} L_{-k\lambda_1} {\bf 1}.
\eeq
The summation set $\Phi(m)$ is the set of all ordered partitions of $m$: $\lambda\in\Phi(m)$ means that $\lambda$ is a multiplet $(\lambda_1,\ldots,\lambda_j)$ for some $j\geq 1$, the only constraints being $\lambda_i\ge1$ for all $i$, and $\sum_{i=1}^j \lambda_i = m$. We denote $|\lambda|=j$ the number of elements in the partition. The coefficients $C_\lambda$ satisfy the recursion relations (recursion on $\sum_i \lambda_i$)
\beq\label{Clambda}
	C_{(\lambda_1,\ldots,\lambda_j)} =
	\delta_{\lambda_j,1} C_{(\lambda_1,\ldots,\lambda_{j-1})}
	+ \sum_{i=1}^j (\lambda_i-1)
	C_{(\lambda_1,\ldots,\lambda_i-1,\ldots,\lambda_j)}
\eeq
for all $j\geq 1$, with initial condition $C_{(1)} = 1$. They can equivalently be obtained from the recursion relations (recursion simultaneously on $\sum_i\lambda_i$ and on $|\lambda|$)
\beq\label{Clambda2}
	C_{(\lambda_1,\ldots,\lambda_j)}
	= \sum_{\ell_i\in\{1,\ldots,\lambda_i\}, i=1,\ldots,j-1
	\atop \ell_j=1} \lt(\sum_{i=1}^j (\lambda_i-\ell_i)\rt)!\;
	\prod_{i=1}^j \frc{(\lambda_i-1)!}{(\lambda_i-\ell_i)!(\ell_i-1)!}
	C_{(\ell_1,\ldots,\ell_{j-1})}
\eeq
for all $j\geq 2$, with initial condition $C_{(n)} = (n-1)!$. We find for instance
\beq\label{Clambda3}
	C_{(1,\ldots,1,\lambda_1,\ldots,\lambda_j)} = C_{(\lambda_1,\ldots,\lambda_j)},\quad
	C_{(n,1,\ldots,1)} = \frc{(n+k-1)!}{k!}
\eeq
where in the second equation, there are exactly $k$ indices that are fixed to 1 on the right of the index $n$. We define similarly the chiral conjugates $\b\cT_{k,m}$.

Using the Virasoro algebra to bring the $L_n$'s with higher $n$ to the right, we have in particular
\beqa
	\cT_{k,1} &=& L_{-k}{\bf 1} \n
	\cT_{k,2} &=& (L_{-k}^2 + (k-1)L_{-2k}) {\bf 1} \n
	\cT_{k,3} &=& (L_{-k}^3 + 3(k-1) L_{-2k}L_{-k} + 2(k-1)(2k-1) L_{-3k}){\bf 1}.
\eeqa
Setting
\beq\label{GCFT}
	{\cal G}_{\rm CFT}:={\rm span}\{{\cal T}_{k,m}:k\geq 2,m\geq1\},\quad
	\b{\cal G}_{\rm CFT}:={\rm span}\{\b{\cal T}_{k,m}:k\geq 2,m\geq1\},
\eeq
we express our main result as follows.
\begin{theorem}\label{res3}
With respect to $\h{\frak{X}}$, the space ${\cal G}_{\rm CLE}$ (\ref{GCLE}) is a reconstruction of ${\cal G}_{\rm CFT}$ (\ref{GCFT}), with bijection linearly generated by
\beq\label{main}
	\tT_{k,m} \equiv {\cal T}_{k,m}.
\eeq
Similarly, the space $\b{\cal G}_{\rm CLE}$ is a reconstruction of $\b{\cal G}_{\rm CFT}$, with bijection generated by $\b\tT_{k,m} \equiv \b{\cal T}_{k,m}$.
\end{theorem}

All these results hold as well in SLE$_{8/3}$.

\begin{rema}
We expect the condition that boundary components of the region be Jordan curves not to be essential, as other finitely connected regions can be obtained by taking appropriate limits.
\end{rema}

\begin{rema}
Evaluating averages of products of identity descendant fields in CFT in simply connected regions is straightforward, using a conformal mapping to the upper half plane, the Cardy boundary conditions on this region, and standard analytic arguments. The insertion of other CFT fields can then be evaluated from their transformation properties. In this case, then, the reconstruction provides explicit results, and the meaning of the reconstruction theorem is clear. For multiply connected regions, however, the situation is more complicated. The problem in CFT can in principle be reduced to the evaluation of the one-point averages of all identity descendants. Indeed, once these are known, the operator algebra can be used to evaluate averages of any products of identity descendants, and the insertion of other CFT fields can be obtained from their transformation properties. The one-point averages of identity descendants depend on the actual conformal boundary conditions, and on the details of the model. The meaning of the reconstruction theorem is that the hypotrochoid fields are part of the right operator algebra, and that insertions of other elements of $\h{\frak X}$ are obtained as in CFT from their transformation properties. This is made precise below by making explicit the underlying Virasoro vertex operator algebra. We do have general expressions for one-point averages, but we have not identified them with one-point averages of identity descendants in CFT, although this identification is fully expected to hold.
\end{rema}

\sect{Analysis and discussion} \label{sectAnalysis}

\subsection{Relation to Virasoro vertex operator algebra through conformal differential operators}

The identification (\ref{main}) with Virasoro descendants imply that expectation values of products of random variables with factors of hypotrochoid fields, $\tT_{k_1,m_1}(w_1)\cdots \tT_{k_p,m_p}(w_p)\;\tX$ where $\tX\in\h{\frak{X}}$, can be evaluated by applying appropriate differential operators on the expectation value of $\tX$. From the CFT point of view this amounts to the conformal Ward identities \cite{BPZ}. In particular, even if $\Or_\tX$ is not a ``known'' CFT field, it is possible to write down the differential operator: we only need to know the conformal transformation properties of $\tX$. As a simple example, if $\tX(x_1,\ldots,x_n)$ is the characteristic function for the event that at least one CLE loop surrounds the points $x_1,\ldots,x_n$ (or winds around them in a specified fashion), then it transforms under a conformal map $g$ as $\tX(x_1,\ldots,x_n)\mapsto g\cdot \tX (x_1,\ldots,x_n) = \tX(g(x_1),\ldots,g(x_n))$. Hence, the insertion of hypotrochoid fields will give rise to conformal Ward identities for a product of $n$ zero-dimension, zero-spin primary fields at positions $x_1,\ldots,x_n$.

More generally, the differential operators are expressed in terms of conformal derivatives, differential operators that act on random variables and on the domain boundary by infinitesimal conformal transformations. The action on random variables implement to the usual Ward identities, and that on the domain boundaries, the boundary conditions (with a light abuse of language, we refer to these together as the conformal Ward identities). On a function $f(\tX,\p A)$, a conformal derivative in the direction of the function $h$, analytic on a domain containing $\p A$ and where $\tX$ is supported, acts as
\beq\label{1storder}
	\nabla_hf(\tX,\p A) = \lt.\frc{d}{dt} f\Big((\id+th)\cdot\tX,(\id+th)(\p A)\Big)\rt|_{t=0},
\eeq
where $g\cdot\tX$ is the conformal transformation of $\tX$ by the conformal map $g$. Note that $h\mapsto \nabla_h f(\tX,\p A)$ is a real linear map in $h$. Holomorphic, resp.~antiholomorphic, conformal derivatives $\Delta[h]$, resp.~$\b\Delta[h]$, are obtained by taking the holomorphic, resp.~antiholomorphic, part in $h$ of the result,
\beq\label{Deltah}
	\Delta[h] = \frc12(\nabla_{h}-i\nabla_{ih}),
	\quad \mbox{resp.} \quad \b\Delta[h] = \frc12(\nabla_{h}+i\nabla_{ih}).
\eeq
Both $\Delta[\cdots]$ and $\b\Delta[\cdots]$ are complex linear maps. Second-order conformal derivatives $\nabla_{h,h'}f(\tX,\p A)$ are obtained by replacing $f(\tX, \p A)$ in (\ref{1storder}) by $\nabla_{h'} f(\tX,\p A)$, seen as a function of $\tX$ and $\p A$; multiple conformal derivatives are obtained recursively.

Note that conformal differentiability and multiple conformal differentiability (and smoothness) requires more than the existence of multiple conformal derivatives with fixed directions $h,h',\ldots$ as above. For instance, $\nabla_{h} f(\tX,\p A)$ is required to be a continuous linear functional of the argument $h$. See \cite{Dcalc,Dhigher} for more details (most of which we will not explicitly need).

In \cite[Thm 4.2]{Dhigher}, the Virasoro vertex operator algebra structure of conformally smooth functions and their conformal derivatives was established. According to this, there is a  linear map ${\bf Y}(\cdot,w)$ from the identity Virasoro module to the space of multiple (holomorphic) conformal derivatives parametrized by $w\in\C$, which allows us to describe the differential operators involved in the conformal Ward identities. The intervening conformal derivatives are conjugated by a function $Z$, which is mainly required to solve a certain infinite set of second order linear conformal differential equations; or equivalently, to transform under conformal maps in a particular way. As mentioned above, in the present context of conformal restriction systems, the independent variables, with respect to which conformal derivatives are taken, are $\tX\in\h{\frak{X}}$ and the region boundary $\p A$. The number $w$ is required to lie in $A$ (there is a unique connected $A$ given $\p A$ in the case of higher connectivity, otherwise $w$ defines $A$ from $\p A$), and $\tX$ is required to be supported in $A$. It turns out that the specific function $Z$ involved only depends on $A$. For this reason, we will denote the map by ${\bf Y}(\cdot,w)_A$.

Let $v$ be a Jordan curve lying in $A$ and surrounding $w$. The linear map ${\bf Y}(\cdot,w)_A$ from the identity module to the space of (conjugated) conformal derivatives can be expressed on basis elements as follows:
\beq\label{Ymap}
	{\bf Y}(L_{\ell_m}\cdots L_{\ell_1}{\bf 1},w)_A =
	Z(\p A,v)^{-1} \Delta[h_{\ell_m,w}]\cdots\Delta[h_{\ell_1,w}] Z(\p A,v)
\eeq
where $\Delta[h_{\ell,w}]$ is a derivative in the direction given by the analytic function
\beq\label{h}
	h_{\ell,w}(z) = -(z-w)^{\ell+1}
\eeq
and where $Z(\p A,v)$ is the {\em relative partition function} (\ref{Zgen}). The operator on the right-hand side of (\ref{Ymap}) is independent of $v$. Using the linear map (\ref{Ymap}), we have, for $w\in A$ and $\tX$ supported in $A\setminus\{w\}$,
\beq\label{eval}
	\mathbb{E}\big[\tT_{k,m}(w)\tX\big]_A = {\bf Y}\lt(\cT_{k,m},w\rt)_A
	\mathbb{E}\big[\tX\big]_A.
\eeq

The relative partition function was defined for simply connected regions in \cite{DTCLE} in the CLE context, and in \cite{Dcalc} in the CFT context (both objects expected to be the same under the identification between CLE and CFT). The main result about it is Theorem \ref{theoZ} (a slight extension and re-writing of \cite[Thm 5.3, 5.5]{DTCLE}), which in particular says that in this case,
\beq\label{DeltaZ}
	\Delta[h_{-2,w}]  Z(\p A,v) = \frc{c}{12} \{s_A,w\}\, Z(\p A,v).
\eeq
where $s$ maps conformally $A$ onto $\uD$ and $\{s,w\}$ is its Schwarzian derivative (again, $v$ must separate $\p A$ from $w$), and $c$ is the central charge. If $A$ is the Riemann sphere (or the plane), then we may take $Z(\emptyset,v)=1$. In the SLE$_{8/3}$ context, since the central charge is zero, we may also take $Z(\p A,v)=1$ for every region $A$. Relation (\ref{DeltaZ}) implies that $\Delta[h_{-2,w}] \log Z(\p A,v)$ transforms in a prescribed way under maps that are conformal on $A$. Theorem \ref{theoZgen} shows that the same transformation property holds in a more general situation where $A$ is finitely connected; this is the transformation property required \cite[Thm 3.2, 4.2]{Dhigher} in order for the map ${\bf Y}(\cdot,w)_A$ above to give rise to the conformal Ward identities. In particular, in the example of the random variable $\tX(x_1,\ldots,x_n)$ above, (\ref{eval}) exactly reproduces the conformal Ward identities for $n$ primary fields of dimension and spin zero.

If $A$ is conformal to the disk, then the operator on the right-hand side of (\ref{Ymap}) can be evaluated explicitly by using the formal relation
\beq
	\Delta[h_{-k,w}] = \frc1{(k-2)!} \lt(\frc{\p}{\p w}\rt)^{k-2}
	\Delta[h_{-2,w}]
\eeq
\cite{Dhigher}, as well as, recursively, (\ref{DeltaZ}). Note that for $g$ conformal on $A$, we may write $s_{g(A)}$ = $s_A\circ g^{-1}$, which gives, using (\ref{1storder}),
\beq
	\Delta[h] \,\p^ns_A(z) = -\p^n (h\,\p s_A(z)).
\eeq

Relation (\ref{eval}) along with the results of \cite{Dhigher} gives us the exact conformal transformation properties of the hypotrochoid fields. Again, these are simply the transformation properties of stress-energy tensor descendants that can be found from standard CFT considerations (see for instance \cite{DFMS97}). Since the conformal derivatives obtained by applying the ${\bf Y}$ map act via conformal transformations of random variables, this means that multiple insertions of hypotrochoid fields can be evaluated by using recursively (\ref{eval}). The result of \cite[Thm 4.2]{Dhigher} is the statement that this reproduces the product of Virasoro vertex operators associated to the descendants $\cT_{k,m}$. Denoting by $Y(\cdot,w)$ the usual Virasoro vertex operator \cite{LL04} (a doubly-infinite formal series in $w$), and choosing the region $A$ to contain the point 0 for convenience, we may express this result as follows:
\beqa\label{voaeval}
	\lefteqn{\Bigg[\mathbb{E}\big[\tT_{k_1,m_1}(w_1)\cdots
	\tT_{k_p,m_p}(w_p)\tX\big]_A\Bigg]_{\mbox{\small Laurent expansion in the region }|w_1|>\ldots>|w_p|}} \\
	&
	\qquad\qquad\qquad\qquad\qquad=& {\bf Y}\lt(Y(\cT_{k_1,m_1},w_1)\cdots
	Y(\cT_{k_p,m_p},w_p) {\bf 1},0\rt)_A
	\mathbb{E}\big[\tX\big]_A.\no
\eeqa
This means that the hypotrochoid fields are part of the correct operator algebra.

\subsection{Consequences of the vertex operator algebra structure}

The vertex operator algebra structure described above mean that hypotrochoid fields satisfy various relations amongst themselves, weakly locally.

Clearly, our main results do not provide a field for every element of the identity module. It is however possible to express $L_{-k}{\bf 1}$ and $L_{-k}^2{\bf 1}$ in terms of hypotrochoid fields:
\beqa\label{Tk1}
	\tT_{k,1}(w) &\equiv& L_{-k}{\bf 1}(w) \\
	\tT_{k,2}(w) - (k-1)\tT_{2k,1}(w) &\equiv& L_{-k}^2{\bf 1}(w).
\eeqa
Higher powers and mixed products of Virasoro generators are involved in linear combinations. Hence, we do not have yet the full vertex operator algebra in terms of fields in conformal restriction systems, so we cannot use the full power of the vertex-algebraic structure. Yet, some aspects of it can be used. A very simple one is the $L_{-1}$-derivative property of vertex operator algebras. This says, for instance, that
\beq
	\frc{d}{dw} L_{-k}{\bf 1}(w) = [L_{-1},L_{-k}]{\bf 1}(w) =
	(k-1)L_{-k-1}{\bf 1}(w)
\eeq
where in the last equality we used the Virasoro algebra. Hence, we obtain
\beq
	\frc{d}{d w} \mathbb{E}\big[\tT_{k,1}(w)\tX\big]
	= (k-1)\mathbb{E}\big[\tT_{k+1,1}(w)\tX\big] 
\eeq
That is, differentiating with respect to $w$ increases the number of arms of the hypotrochoid by one.

Using derivatives of $\tT_{k,2}(w)$ we can in fact obtain expressions for all descendants involving two $L$-operators. Indeed, we have
\[
	\frc{d^n}{dw^n} \tT_{k,2}(w) \equiv
	\Big[\lt({\rm ad}L_{-1}\rt)^n(L_{-k}^2) + (k-1)(2k-1)2k\cdots (2k+n-2)
	L_{-2k-n}\Big]{\bf 1}(w)
\]
and $\lt({\rm ad}L_{-1}\rt)^n(L_{-k}^2)$ involves products $L_{-k}L_{-k'}$ with $|k-k'|\leq n$. Using the Virasoro algebra, one can then solve recursively for $L_{-k}L_{-k'}$ for $k'<k$. For instance,
\beqa
	\lefteqn{\frc1{2(k-1)}\lt(\frc{d}{dw} \tT_{k,2}(w) - 2k(k-1)\,\tT_{2k+1,1}(w)\rt)} &&\n
	&\hspace{5cm}\equiv& L_{-k-1}L_{-k}{\bf 1}(w)\hspace{5cm}\n
	\lefteqn{\frc1{2(k-1)k}\lt(\frc{d^2}{dw^2} \tT_{k,2}(w)
	- 2(k-1)\,\tT_{k+1,2}(w) + 2k(k-1)(2k-1)\,\tT_{2k+2,1}(w) \rt)} &&\n
	&\hspace{5cm}\equiv& L_{-k-2}L_{-k}{\bf 1}(w).
\eeqa
This implies that we can also express all $L_{-k}^3{\bf 1}(w)$ in terms of hypotrochoid fields and their derivatives, for instance
\beq
	\tT_{2,3}(w) - \frc34 \frc{d^2}{dw^2} \tT_{2,2}(w) + \frc32 \tT_{3,2}(w) + 3\,\tT_{6,1}(w) \equiv L_{-2}^3{\bf 1}(w).
\eeq

Probably the deepest consequence of the vertex operator algebra structure is that of the operator product expansion. The result (\ref{voaeval}) mean that in any expectation value, the product of two hypotrochoid fields can be expressed as a series expansion in integer powers of their distance, with coefficients that are other local variables corresponding to descendants of the identity. Since we do not have expressions for all descendants in terms of hypotrochoid fields, we do not have the full operator product expansion in general -- the hypotrochoid fields do not form a close operator algebra. However, we do have, in principle, the operator product expansions (OPEs) for products $\tT_{k,1}(w)\tT_{k',1}(w')$: from (\ref{Tk1}),  CFT results \cite{DFMS97} imply that these OPEs only involve the fields associated to $L_{-k}{\bf 1}(w)$ and those associated to $L_{-k}L_{-k'}{\bf 1}(w)$, hence only $\tT_{k,1}(w)$ and $\tT_{k,2}(w)$ and their derivatives.

\sect{Conformal restriction system}\label{sectCRS}

\subsection{Definition} \label{ssectDefCR}

We understand a region as an open, connected and finitely connected subset of the Riemann sphere $\hC=\C\cup\{\infty\}$, possibly $\hC$ itself. For regions that are proper subsets of $\hC$, we will restrict ourselves to regions whose boundary components are Jordan curves (whence the following conformal restriction system imposes weaker conditions).

Let $\frak{X}$ be an associative, commutative algebra over $\C$ with a unit ${\bf 1}$. To every element $\tX\in\frak{X}$, we associate a set $\Supp(\tX)$ of closed subsets of $\hC$, the supports of $\tX$, with the following properties:
\bi
\item $\emptyset\in\Supp({\bf 1})$
\item $K\in\Supp(\tX)\Rightarrow J\in\Supp(\tX)$ for every closed sets $J\supset K$ and every $\tX\in\frak{X}$
\item $\Supp(a\tX) = \Supp(\tX)$, $\Supp(\tX+\tY) \supset \Supp(\tX)\vee \Supp(\tY)$ and $\Supp(\tX\tY) \supset \Supp(\tX)\vee \Supp(\tY)$ for every $\tX,\tY\in\frak{X}$, $a\in\C$, where $\Supp(\tX)\vee \Supp(\tY):= \{J\cup K:J\in\Supp(\tX), K\in \Supp(\tY)\}$.
\ei
We then say that $\frak{X}$ is a supported algebra (if we do not have the algebra structure, a supported linear space). Given a region $A$, there is an associated subalgebra $\frak{X}^{(A)}\subset \frak{X}$ of elements supported in $A$ (that is, that possess at least one support not intersecting $\hC\setminus A$). For every region $A$, let $\mathbb{E}\big[\cdot\big]_A$ be a linear functional on $\frak{X}^{(A)}$ with $\mathbb{E}\big[{\bf 1}\big]_{A} = 1$.

As in Subsection \ref{ssectCLECFT}, we may provide a meaning for convergent sequences in $\frak{X}$ using the weak-local topology associated to the set of linear functionals $\{\mathbb{E}\big[\cdot\big]_A:\mbox{regions $A$}\}$ (Definition \ref{defiwl}). Again as explained there, there is a natural meaning for the resulting limit of such a sequence, which we may refer to as a ``renormalized element'', with its associated supports. Thus we can complete the supported linear space $\frak{X}$ with respect to this topology, giving a supported linear space $\cl{\frak{X}}$ equipped with a set of linear functionals as above. Note that $\cl{\frak{X}}$ is not, in general, an algebra. Let us see $\frak{X}$ as a quasi-algebra: an algebra where the product $\tX\tY$ is defined whenever there exists a support of $\tX$ and a support of $\tY$ that are disjoint. If ${\cal G}$ is a linear space that lies in $\cl{\frak{X}}\setminus \frak{X}$ and that is consistent (Definition \ref{defics}), then, as in Subsection \ref{ssectCLECFT}, we may extend the supported quasi-algebra $\frak{X}$ to the supported quasi-algebra $\frak{X}_{\cal G}$ by linearly adjoining products of elements in ${\cal G}$ that possess disjoint supports, and $\frak{X}_{\cal G}$ is also equipped with a set of linear functionals as above.

We say that the set of subalgebras $\{\frak{X}^{(A)}:\mbox{regions $A$}\}$ forms a linear representation of the groupoid of (univalent) conformal maps if the following holds: For every region $A$ and for every conformal map $g$ on $A$ there is an algebra isomorphism
\beqa
	\frak{X}^{(A)}&\to&\frak{X}^{g(A)}\n
	\tX&\mapsto& g\cdot\tX \no
\eeqa
that agrees with compositions of conformal maps: $g\cdot(g'\cdot \tX) = (g\circ g')\cdot \tX$ whenever this makes sense. This definition can immediately be adapted to supported quasi-algebras and linear spaces.

We say that $N\subset \hC$ is a tubular neighborhood if it is the closure of an annular region with Jordan boundaries. Let $\alpha$ be a Jordan curve. If $\alpha\subset N$ and $\alpha$ separates the two components of $\hC\setminus N$, then we say that $N$ is a tubular neighborhood of $\alpha$. We say that a sequence of tubular neighborhoods $N_j$ of $\alpha$ converges to $\alpha$ if for every region $A$ containing $\alpha$, there is a $k$ such that for all $j>k$, $N_j\subset A$. Note that this is equivalent to convergence in the Hausdorff topology. Let $f$ be a complex set function. We say that $\lim_{N\to\alpha} f(N)$ exists and equals $a\in\C$ if for every sequence $N_j$ of tubular neighborhoods of $\alpha$ converging to $\alpha$, $\lim_{j\to\infty} f(N_j)$ exists and gives $a$.

A {\em conformal restriction system} is a supported algebra $\frak{X}$ equipped with a set of linear functionals $\{\mathbb{E}\big[\cdot\big]_A:\mbox{regions $A$}\}$ with $\mathbb{E}\big[{\bf 1}\big]_{A} = 1$, and a linear representation on $\frak{X}$ of the groupoid of conformal maps, with the following properties:
\newcounter{savez}
\begin{enumerate}
\item For every tubular neighborhood $N$ there exists $\tI(N)\in\frak{X}$ such that $N\in \Supp(\tI(N))$ and $g\cdot\tI(N) = \tI(g(N))$, and satisfying the condition that $\mathbb{E}\big[\tI(N)\big]_A \neq 0$ for every $N$ and every $A\supset N$.
\setcounter{savez}{\value{enumi}}
\end{enumerate}
Let us define
\beq\label{tEN}
	\tE(N):=\frc{\tI(N)}{\mathbb{E}\big[\tI(N)\big]_\hC}.
\eeq
We have in particular
\beq\label{gEN}
	g\cdot\tE(N) = \frc{\mathbb{E}\big[\tI(g(N))\big]_\hC}{\mathbb{E}\big[\tI(N)\big]_\hC} \tE(g(N)).
\eeq
\begin{enumerate}
\setcounter{enumi}{\value{savez}}
\item The limit
\beq\label{tEalpha}
	\tE(\alpha):=\lim_{N\to\alpha} \tE(N)
\eeq
exists and is nonzero weakly locally (Definition \ref{defiwl}) with respect to $\frak{X}$ for every $\alpha$. Note that $\alpha\in\Supp(\tE(\alpha))$.
\item The set ${\cal G}:=\{\tE(\alpha):\mbox{Jordan curves $\alpha$}\}$ forms a consistent set (Definition \ref{defics}) with respect to $\frak{X}$. We have the corresponding quasi-algebra $\h{\frak{X}}:=\frak{X}_{\cal G}$.
\item {\bf Conformal invariance.} We have $\mathbb{E}\big[g\cdot \tX\big]_{g(A)} = \mathbb{E}\big[\tX\big]_{A}$ for every $\tX\in\frak{X}$ supported in the region $A$ and every map $g$ conformal on $A$.
\setcounter{savez}{\value{enumi}}
\end{enumerate}
The latter point along with Points 1 and 2 in particular implies that the limit
\beq\label{F}
	F(g,\alpha) := \lim_{N\to\alpha} \frc{\mathbb{E}\big[\tI(N)\big]_\hC}{\mathbb{E}\big[\tI(g(N))\big]_\hC}
\eeq
exists, and that we can define
\beq\label{transfoE}
	g\cdot \tE(\alpha) = \frc{\tE(g(\alpha))}{F(g,\alpha)},
\eeq
giving the quasi-algebra $\h{\frak{X}}$ the structure of a linear representation of the groupoid of conformal maps, and extending Conformal invariance to $\h{\frak{X}}$.
\begin{enumerate}
\setcounter{enumi}{\value{savez}}
\item {\bf Restriction.} Let $\alpha$ be a Jordan curve lying in the region $A$, and let $\tX\in\h{\frak{X}}$ be a product of factors each supported in some connected component of $A\setminus \alpha$ (which may be different for the different factors). Then
\beq\label{restr}
	\frc{\mathbb{E}\big[\tE(\alpha)\,\tX\big]_{A}}{
	\mathbb{E}\big[\tE(\alpha)\big]_{A}
	}
	=\mathbb{E}\big[\tX\big]_{A\setminus \alpha}
\eeq
where by definition $\mathbb{E}\big[\tY\tZ\big]_{B\cup C}$ = $\mathbb{E}\big[\tY\big]_{B}\mathbb{E}\big[\tZ\big]_{C}$ whenever $B$ and $C$ are disjoint regions and $\tY$, $\tZ$ are supported in $B$, $C$ respectively.
\item {\bf Smoothness.} Let $A$ be a region and $\tX\in\h{\frak{X}}$ supported in $A$. Let $A_i$ be simply connected regions such that $A=\cap_i A_i$ (these always exist). If $g$ is a univalent conformal map on an open neighborhood $B$ of $\p A_i$, let $g\cdot A_i$ be the simply connected region bounded by $g(\p A_i)$ such that $g(A_i\cap B)\subset g\cdot A_i$. For every set $I$ of indices $i$, denote by $g_I\cdot A := {\cap_{i\in I} (g\cdot A_i)\cap (\cap_{i\not\in I}A_i})$. Let $U$ be a closed annular set containing either (i) $K\cup \cup_{i\in I}\p A_i$ for some $K\in\Supp(\tX)$, or (ii) only $\cup_{i\in I}\p A_i$. Then for every $I$, the function (i) $g\mapsto \mathbb{E}\big[g\cdot \tX\big]_{g_I\cdot A}$, or (ii) $g\mapsto \mathbb{E}\big[\tX\big]_{g_I\cdot A}$, respectively, is smooth at the identity with respect to conformal maps $g$ on $U$, in the sense of \cite{Dhigher}. We will say that we are taking conformal derivatives with respect to $\tX$ and/or $\cup_{i\in I}\p A_i$.

\item {\bf Local covering.} For $\tX$ and $A$ as in the above point, let $\nabla$ be a conformal derivative symbol, of any order, with respect to $\tX$ and/or $\cup_{i\in I}\p_i A$ (including $\nabla=1$). Let $K_\ep\in A:\ep>0$ be a family of closed, connected sets in the region $A$. Let $D_\ep$ be the smallest closed disk covering $K_\ep$. Assume that $\lim_{\eta\to0} \cup_{\ep\in(0,\eta)} D_\ep = z\in A$ (that is, intuitively, the family of sets $K_\ep$ tends to the point $z$ as $\ep\to0$). Then
\beq
	\lim_{\ep\to0}
	\nabla\mathbb{E}\big[\tX\big]_{A\setminus K_\ep}
	=\nabla\mathbb{E}\big[\tX\big]_{A}
\eeq
for every $\tX\in\h{\frak{X}}$ supported in $A\setminus \{z\}$.
\end{enumerate}
This system is expressed here in terms of linear functionals $\mathbb{E}\big[\cdot\big]_A$ on general regions $A$, but it is clear that it can be restricted to regions whose boundary components, if any, are Jordan curves. In particular, in Point 6, for every $g$ conformal on $K$ near enough to the identity in the topology of \cite{Dhigher}, we have that $g_I\cdot A$ has Jordan boundary components if $A$ does. Also, in Point 7, we may restrict to closed sets $K_\ep$ such that $A\setminus K_\ep$ has Jordan boundary components for every $\ep>0$.

Note that by definition, the factors $F(g,\alpha)$ satisfy the composition rule
\beq\label{compoF}
	F(g\circ g',\alpha) = F(g,g'(\alpha)) F(g',\alpha)
\eeq
for every conformal map $g'$ on a neighborhood of $\alpha$, and $g$ on a neighborhood of $g'(\alpha)$. Further, by conformal invariance,
\beq\label{Mobinv}
	F(G,\alpha) = 1
\eeq
for every M\"obius map $G$.

Note also that a simple consequence of multiple applications of the Restriction property is the following. Let $\{\alpha_i\}$ be a finite set of disjoint Jordan curves all lying in the region $A$, and let $\tX\in\h{\frak{X}}$ be a product of factors each supported on some connected component of $A\setminus \cup_i\alpha_i$. Then
\beq\label{mulrest}
	\mathbb{E}\big[\tX\big]_{A\setminus \cup_i\alpha_i}
	=
	\frc{\mathbb{E}\Big[\prod_i \tE(\alpha_i)\,\tX\Big]_{A}}{
	\mathbb{E}\Big[\prod_i\tE(\alpha_i)\Big]_{A}
	}.
\eeq

Finally, sometimes it is convenient to assume the presence of a ``complex conjugation'' involution in the algebra, $\tX\mapsto \b{\tX}$, with the usual properties.

\begin{rema}
It is interesting to note the structural similarity between our conformal restriction system and aspects of algebraic quantum field theory. We believe that the present notions could be useful in developing a formulation paralleling that of algebraic QFT, but in the context of statistical systems on Euclidean space rather than quantum systems on Minkowski space.
\end{rema}

\subsection{Expected relation to CLE}

Here is an example of a supported algebra. Consider a loop configuration on a region $A$ to be a set of Jordan curves all lying in $A$. Consider further the set of all loop configurations on the Riemann sphere $\hC$. Then we may take $\frak{X}$ to be the algebra of complex functions on the set of loop configurations on $\hC$, with support defined as in Subsection \ref{ssectCLECFT}, and $\mathbb{E}\big[\cdot\big]_A$ to be the expectation functional associated to a measure on loop configurations on $A$. We may also take $\tI(N)$ to be the function that takes the value one if there is at least one loop $\gamma$ such that $N$ is a tubular neighborhood of $\gamma$, and zero otherwise.

This example naturally connects with CLE. In fact, we expect the CLE measures in the dilute regime (here, assumed to exist on every region $A$) to give rise to a conformal restriction system, by identifying $\mathbb{E}$ with the CLE expectation value, and by choosing an appropriate choice of algebra $\frak{X}$ of CLE random variables containing all variables $\tI(N)$ as defined in the above paragraph. Of course, in the dilute CLE case, the loops in the configurations satisfy almost surely some additional properties: they are disjoint, there are countably-many of them, for any $r>0$, the number of loops of diameter larger than $r$ is finite, and almost every point is surrounded by at least one loop (in fact, by infinitely-many loops).

We justify in CLE some of the points of a conformal restriction system as follows, following the ideas in \cite[Sect 3]{DTCLE}. Point 1 is quite immediate. Points 2 and 3, on the other hand, are rather delicate. Let us consider only Point 2. Let us denote by $P_A(\cdot)$ the CLE probability function on the region $A$, and by $\ev(N)$ the event for which $\tI(N)$ is the indicator variable. For (appropriate) events $\tou$, the existence of the limits
\[
	\lim_{N\to\alpha} \frc{P_A(\tou\cap \ev(N))}{P_A(\ev(N))} = \lim_{N\to\alpha} P_A(\tou|\ev(N))
\]
is immediately expected simply from the nesting property of CLE. But Point 2 is more general: in the probability notation, it implies that $\lim_{N\to\alpha} P_A(\tou\cap \ev(N))/P_B(\ev(N))$ exists for different regions $A$ and $B$.

If Points 2 and 3 can be established in CLE, Conformal invariance and Restriction are very natural. Conformal invariance, in particular with (\ref{gEN}), is a fundamental property of CLE on any region. Concerning Restriction, note that the nesting property of CLE referred to above should give us more than the existence of the limit $\lim_{N\to\alpha} P_A(\tou|\ev(N))$, but also should tell us that this equals $P_C(\tou)$ where $C$ is the component of $A\setminus \alpha$ containing the support of $\tou$. This holds at least whenever $A$ is simply connected and $\tou$ is supported on the simply connected component of $A\setminus\alpha$. Indeed, in this case, the event $\ev(N)$ guarantees, in the limit $N\to\alpha$, that, informally, a loop takes the shape $\alpha$, and nesting says that inside this loop, we find a CLE measure on the domain it bounds.

Smoothness is much harder to argue for, but entirely expected: expectation values should lead to functions that are smooth under any small deformations of domain boundaries and random variables. This should hold {\em a fortiori} under small conformal deformations.

Finally, Local covering is expected in CLE from the almost-sure existence of infinitely many loops surrounding any point. As the set $K_\ep$ approaches the point $z$ as described above, there will be more and more loops surrounding it. We may take these loops not to intersect some support of $\tX$, and in the limit, where infinitely many loops separate this support from $K_\ep$, there should be factorization. Note that in CLE, the regions $A\setminus\{z\}$ and $A$ are expected to give rise exactly to the same measure.

It is in fact possible to modify the relation between CLE and conformal restriction system by using, for the event $\ev(N)$ (associated to the variable $\tI(N)$), that according to which no loop intersect both components of $\hC\setminus N$ simultaneously. In this case, the CLE justification of Restriction, for instance, uses the probabilistic conformal restriction property of CLE, instead of nesting.

Note that in CLE, the limit $\lim_{N\to\alpha} P_A(\ev(N))$ (in both definitions of $\ev(N)$) is expected to vanish. Hence, it is indeed necessary to defined renormalized random variables by a limit as in Point 2 above.

\subsection{Some implications} \label{ssectImplic}

Let $C,D$ simply connected Jordan domains with $\cl{D}\subset C$. Then we define the relative partition function following the definition in the CLE context \cite[Def 2.3]{DTCLE} (re-written in the present notation):
\beq\label{Z}
	Z(\p C, \p D) := \frc1{
	\mathbb{E}\big[\tE(\p C)\big]_{\hC\setminus \cl D}}.
\eeq
We also admit $C=\hC$, in which case $\p C= \emptyset$ and $Z(\emptyset, \p D) = 1$. It is straightforward to see that, thanks to (\ref{Mobinv}) and (\ref{transfoE}), $Z(G(\p C),G(\p D)) = Z(\p C,\p D)$ for every M\"obius map $G$. We have the following result.
\begin{theorem}[\cite{DTCLE} and forthcoming work]\label{theoZ}
Let $u=\p C,v=\p D$ be disjoint Jordan curves. Then $Z(u,v) = Z(v,u)$. Let $C$ and $D$ be the Jordan domains bounded by $u$ and $v$, respectively, such that $\cl D\subset C$. Then there exists a complex number $c$ such that, for every $w\in D$, $w\neq\infty$,
\beq\label{DeltaZtheo}
	\Delta[h_{-2,w}] \log Z(\p C,\p D) = \frc{c}{12} \{s,w\}
\eeq
where $s$ maps conformally $C$ onto the unit disk $\uD$, and where the holomorphic conformal derivative $\Delta[h_{-2,w}]$ is with respect to conformal transformations of the set $\p C\cup \p D$.
\end{theorem}
In the context of CLE, a version of Theorem \ref{theoZ}, where $u,v$ are required to be ``smooth enough'' (for technical reasons -- see \cite{DTCLE}), is a consequence of the results of \cite{DTCLE}, in particular \cite[Thm 5.3, 5.5]{DTCLE}. In \cite{DTCLE}, proofs were provided based on some of the basic properties of CLE (including conformal invariance), as well as four hypotheses, \cite[Hyp 3.1, 3.2, 3.3, 5.1]{DTCLE}. We show in the next subsection that the four hypotheses are consequences of the present conformal restriction system. This implies that the results of \cite{DTCLE}, including the ``smooth'' version of the theorem above, are results holding in the context of conformal restriction systems. We will provide, in a forthcoming work, an independent proof of Theorem \ref{theoZ} in the context of conformal restriction systems, and without the restriction on $u$ and $v$ being smooth enough.

\begin{rema}
Theorem \ref{theoZ} indicates, in particular, that to every conformal restriction system there is an associated {\em central charge} $c$.
\end{rema}

\begin{rema}
From (\ref{DeltaZtheo}), we see that $\Delta[h_{-2,w}] \log Z(\p C,\p D)$ is in fact independent of $\p D$, as long as $w\in D$. We may wish then to take the limit where, either, $\p D\to\p C$, or where $D\to w$ (in an appropriate fashion). However, no axiom of the conformal restriction system indicates any simplification arising in such limits. In particular, note that although Local covering implies that $\lim_{D\to w} Z(\p C,\p D) = 1$, it does not imply $\lim_{D\to w} \Delta[h_{-2,w}] \log Z(\p C,\p D) = 0$, because the derivative $\Delta[h_{-2,w}]$ is with respect to conformal transformations of $\p C\cup \p D$, not just $\p C$. It may be that the operations $\lim_{D\to w}$ and $\Delta[h_{-2,w}]$ can be interchanged in a special conformal restriction system; in this case, it would be a system with zero central charge, $c=0$.
\end{rema}

Before showing that the hypotheses of \cite{DTCLE} hold in conformal restriction systems, we first generalize the above considerations to finitely connected regions. Let $C$ be a finitely connected region whose boundary components, if any, are Jordan curves. We can always write $C = \hC\cap \cap_i C_i$ where $C_i$ (if any) are Jordan domains with pairwise disjoint complements. For definiteness, we take the index $i$ to start from 1 and end at $n$. It will be natural in Section \ref{sectProofs} to define a relative partition function for such general regions $C$ as follows: given a Jordan domain $D$ such that $\cl D\subset C$,
\beq\label{Zgen}
	Z(\p C,\p D) := \frc{
	\mathbb{E}\big[\prod_{i=1}^n \tE(\p C_i)\big]_{\hC}}{
	\mathbb{E}\big[\prod_{i=1}^n \tE(\p C_i)\big]_{\hC\setminus \cl D}}.
\eeq
Then a quite surprising consequence of Theorem \ref{theoZ}, of Restriction, and of basic results in the theories of conformal derivatives and of conformal maps, is the following very general result:
\begin{theorem} \label{theoZgen} Let $C$ and $D$ be as above, and $w\in D$, $w\neq \infty$. Then, for every map $g$ univalent conformal on $C$,
\beq\label{DeltaZtransfo}
	(\p g(w))^2 \lt(\Delta[h_{-2,g(w)}] \log Z\rt)(g(\p C),g(\p D)) +
	\frc{c}{12} \{g,w\} = \Delta[h_{-2,g(w)}] \log Z(\p C,\p D)
\eeq
where $c$ is the complex number involved in Theorem \ref{theoZ}.
\end{theorem}
\proof Thanks to (\ref{DeltaZtheo}), one can easily check that this holds whenever $C$ is simply connected (case $n=1$). For the multiply connected case ($n>1$), by Restriction, we have
\beq
	Z(\p C,\p D) = Z(\p C_1,\p D) \t{Z}(\{\p C_i\},\p D),\quad
	\t{Z}(\{\p C_i\},\p D) := \frc{
	\mathbb{E}\big[\prod_{i=2}^n \tE(\p C_i)\big]_{C_1}}{
	\mathbb{E}\big[\prod_{i=2}^n \tE(\p C_i)\big]_{C_1\setminus \cl D}}.
\eeq
Thanks to (\ref{transfoE}), the factor $\t{Z}(\{\p C_i\},\p D)$ is invariant under transformations that are univalent conformal on $C_1$. Hence, by \cite[Prop 3.9]{Dhigher} (or \cite[Cor 3.11]{Dcalc}), we have
\beq
	(\p g(w))^2 \lt(\Delta[h_{-2,g(w)}] \log \t{Z}\rt)(\{g(\p C_i)\},g(\p D))
	= \Delta[h_{-2,g(w)}] \log \t{Z}(\{\p C_1\},\p D).
\eeq
This along with the fact that (\ref{DeltaZtransfo}) holds for $\log Z(\p C_1,\p D)$ (simply connected case) implies (\ref{DeltaZtransfo}) for every map $g$ univalent conformal on $C_1$. The choice of $C_1$ is of course arbitrary, hence this holds with $C_1$ replaced by any $C_i$. The generalization to every $g$ univalent conformal on $C$ is obtained thanks to the factorization theorem for conformal maps on finitely connected regions \cite{fact1,fact2,Dfact}. This theorems sttes that any such $g$ may be written as the composition $g=g_1\circ\cdots\circ g_n$ where $g_i$ are conformal on simply connected regions. Hence by recursive use of (\ref{DeltaZtransfo}) with $g$ conformal on simply connected regions, and by the fact that the transformation property (\ref{DeltaZtransfo}) agrees with compositions of conformal maps, we obtain (\ref{DeltaZtransfo}) in the general case.
\eproof

\begin{rema}
In the above corollary, we restrict ourselves to maps $g$ that are conformal on $C_1$. This is sufficient for our present purposes. However, we can obtain a formula in the more general situation where $g$ is conformal on $C$ by using recursively (\ref{DeltaZtransfo}), from the fact that any map $g$ conformal on $C$ is a composition of maps conformal on simply connected domains. We will come back to this in a forthcoming work.
\end{rema}

\subsection{Relation with \cite{DTCLE}}

We now verify that the hypotheses of \cite{DTCLE} hold in conformal restriction systems.

First, Hypothesis \cite[Hyp 3.1]{DTCLE} is an immediate consequence of Local covering, and Hypothesis \cite[Hyp 3.2]{DTCLE} is an immediate consequence of Point 2 and of Restriction.

We may show a slightly more general version of the first part of Hypothesis \cite[Hyp 3.3]{DTCLE}, using Points 2 and 3, and especially Restriction. Let $C\subset \hC$ be a region, $A$ be a Jordan domain, and $B$ be a region with Jordan boundaries $\p_i B$, such that $\cl{A}\subset B$ and $\cl{B}\subset C$. Then thanks to (\ref{mulrest}),
\beqa
	\lim_{N\to \p A} \frc{P_B(\ev(N))}{P_C(\ev(N))} &=&
	\frc{\mathbb{E}\big[\tE(\p A)\big]_B}{
	\mathbb{E}\big[\tE(\p A)\big]_C } \n
	&=& 	\frc{\mathbb{E}\big[\tE(\p A)\prod_i \tE(\p_i B)\big]_C}{
	\mathbb{E}\big[\tE(\p A)\big]_C
	\mathbb{E}\big[\prod_i\tE(\p_i B)\big]_{C} } \n
	&=& 	\frc{\mathbb{E}\big[\prod_i \tE(\p_i B)\big]_{C\setminus\cl A}}{
	\mathbb{E}\big[\prod_i\tE(\p_i B)\big]_{C}} =
		\lim_{\{N_i\to \p_i B\}} \frc{
	P_{C\setminus \cl A}\lt(\prod_i\ev(N_i)\rt)}{
	P_{C}(\prod_i \ev(N_i))}\label{relaa}
\eeqa
The second part of Hypothesis \cite[Hyp 3.3]{DTCLE} is a consequence of a similar calculation, along with Local covering.

Hypothesis \cite[Hyp 5.1]{DTCLE} concerned continuous differentiability of certain CLE probabilities. It involved certain new renormalized variables related to $\tE(\alpha)$. In our present derivation, we do not need these variables, but for completeness and in order to guarantee that the results of \cite{DTCLE} hold in a conformal restriction system, we discuss them now, in the present notation. For technical reasons we restrict ourselves to a subset of Jordan curves that are ``smooth enough'' \cite[Sect 4]{DTCLE}. Let $\dom$ be the set of Jordan domains $A$ such that any conformal map $g:\uD\tto A$ can be extended to a conformal map on a neighborhood of $\cl\uD$. Then also any conformal map $g':\hC\setminus \cl \uD \tto \hC\setminus \cl A$ has a conformal extension to a neighborhood of $\hC\setminus \uD$ (in particular, if $A\in\dom$ then $\hC\setminus \cl A\in\dom$). In \cite[Sect 4]{DTCLE}, a procedure was set-up, for every $A\in\dom$, for defining a M\"obius invariant ``renormalized weight'' associated to the event that there be at least one CLE loop of the shape $\p A$. First the regularized events $\ev_\eta(A):=\ev(\cl{A}\setminus A_\eta)$, $\eta>0$ are defined, where $A_\eta\subset A$ is another Jordan domain. The domain $A_\eta$ is defined by choosing, for every $A$, a conformal map $g_A:\hC\setminus \uD\tto \hC\setminus A$, and by setting
\beq\label{Aeta}
	\hC\setminus A_\eta := g_A \lt(\hC\setminus (1-\eta)\uD\rt).
\eeq
The only requirement on $g_A$ is that if two domains $A$ and $A'$ are related to each other by a M\"obius map, then so should be $g_A$ and $g_{A'}$. This requirement can be solved, as the construction of \cite[Sect 4]{DTCLE} shows; in particular, M\"obius transformations produce a fibration of $\Upsilon$, and we have to choose a section $\Omega$ along with maps $g_A$ for all $A\in\Omega$. Note that the construction guarantees that if $C\supset \hC\setminus A$ is a region where $g_A^{-1}$ is conformal, then
\beq\label{relzz}
	P_C(\ev_\eta(A)) = P_{g_A^{-1}(C)}(\ev_\eta(\uD)).
\eeq

The construction of \cite[Sect 4]{DTCLE} provides a renormalized random variable defined as
\beq\label{defU}
	\tU(A):=\lim_{\eta\to0} \frc{1\lt[\ev_\eta(A)\rt]}{P_{\hC}(\ev_\eta(\uD))}
\eeq
for every $A\in\dom$. That is, in contrast with (\ref{tEalpha}), the way the limit is taken is more prescribed, and the denominator is independent of $A$. This is a well-defined renormalized variable (that is, the limit exists weakly locally), simply related to $\tE(\p A)$. Indeed, let $C\supset \hC\setminus A$ be a region where $g_A^{-1}$ is conformal. Then, thanks to (\ref{relzz}), we have
\beq\label{tUtE}
	\tU(A) = \tE(\p A)\,\frc{\mathbb{E}\big[\tE(\p\uD)]_{g_A^{-1}(C)}}{
	\mathbb{E}\big[\tE(\p A)\big]_C}.
\eeq
Note that the fact that the right-hand side is independent of $C$ is nontrivial. Clearly, the renormalized random variable $\tU(A)$ has $\p A$ as a support.

Theorem \cite[Thm 4.1]{DTCLE} then shows that the renormalized random variable $\tU(A)$ transforms as
\beq\label{deff}
	g\cdot \tU(A) = \frc{\tU(g(A))}{f(g,A)}
\eeq
for any map $g$ conformal on $A$, where $f(g,A)$ is a real positive factor. Further, the factor $f(g,A)$ is equal to 1 if $g$ is a M\"obius map; this means that the renormalization process described by (\ref{defU}) preserves M\"obius invariance. These results are quite immediate in our present conformal restriction system: from (\ref{transfoE}) and (\ref{tUtE}) we find
\beq
	f(g,A) = \frc{F(g\circ g_A,\p \uD)}{F(g_{g(A)},\uD)}.
\eeq
In particular, it is obvious that $f$ is not identically equal to one because of the ambiguity in defining $g_A$, and that for $g$ a M\"obius transformation it is one by the requirements of our construction. By definition, the function $f$ satisfies a composition relation like (\ref{compoF}):
\beq
	f(g\circ g',A) = f(g,g'(A)) f(g',A).
\eeq

In terms of the variables $\tU(N)$, Relations (\ref{relaa}) and (\ref{tUtE}) yield, in the simply connected case,
\beq
	\frc{\mathbb{E}\big[\tU(A)\big]_B}{\mathbb{E}\big[\tU(A)\big]_C}
	=\frc{
	\mathbb{E}\big[
	\tU(\hC\setminus B)\big]_{C\setminus \cl A}}{
	\mathbb{E}\big[\tU(\hC\setminus B)\big]_{C}},
\eeq
which is the crucial relation used in \cite{DTCLE}.

Finally, Hypothesis \cite[Hyp 5.1]{DTCLE} is a consequence of Smoothness. More precisely, the first part follows from Smoothness and the explicit form (\ref{tUtE}) for the variable $\tU(A)$, and the last part directly follows from our general expression for Local covering, including the general multiple conformal derivatives operator $\nabla$.

\sect{Proofs} \label{sectProofs}

\subsection{Conformal derivatives}

The key results used in establishing the reconstruction described in Section \ref{sectResults} are those of the works \cite{Dhigher}, where the conformal Ward identities associated to all descendants of the identity field are expressed in terms of conformal derivatives. In order to establish the reconstruction, we need some crucial technical lemmas about conformal derivatives.

Let $h$ be holomorphic on an annular domain $C$ and let
\beq\label{gep}
	g_\ep = \id + \ep h.
\eeq
There exists a compact annular subset $K\subset C$ and an open neighborhood $N\subset \R$ of $0$ such that for every $\ep\in N$, both $g_\ep$ and its inverse $g_\ep^{-1}$ are conformal on $K$. Consider the set $\Omega$ of conformal maps on $K$, with the topology defined in \cite{Dhigher}. Then there is an open subset ${\cal N}\subset\Omega$ that contains the identity, such that $g_\ep,\,g_\ep^{-1}\in{\cal N}$ for all $\ep\in N$.
\begin{lemma}\label{lemexpf}
Let $g_\ep$ and ${\cal N}$ be as above. Consider a function $f$ that is smooth on ${\cal N}$. Then $f(g_\ep)$ and $f(g_\ep^{-1})$ have asymptotic expansions in nonnegative powers of $\ep$,
\beq\label{expgeq}
	f(g_\ep) \sim \sum_{m\geq 0} \frc{\ep^m}{m!} \,\sqa_h^{(m)}f(\id),\quad
	f(g_\ep^{-1}) \sim \sum_{m\geq 0} \frc{\ep^m}{m!} \,\t\sqa_h^{(m)}f(\id)
\eeq
where $\sqa_h^{(m)}$ and $\t\sqa_h^{(m)}$ are both $m$-linear in $h$ (that is, for instance, $\sqa_{ah}^{(m)} f(\id) = a^m\sqa_h^{(m)}f(\id)$ for every $a\in\R$), and where $\sqa_h^{(0)} = \t\sqa_h^{(0)} = 1$ and $\sqa_h^{(1)} = -\t\sqa_h^{(1)} = \nabla_h$. Further, the conformal differential operators $\sqa_h^{(m)}$ are given, for $m\geq 2$, in terms of multilinear intermediate operators $\sqa_{h_m,\ldots,h_1}^{(m)}$ by
\beq\label{sqa1}
	\sqa^{(m)}_h = \sqa^{(m)}_{h,h,\ldots,h}
\eeq
and these intermediate operators satisfy the recursion relations
 \beq\label{sqa2}
	\sqa^{(m)}_{h_m,\ldots,h_1} = \nabla_{h_m} \sqa^{(m-1)}_{h_{m-1},\ldots,h_1} - \sum_{j=1}^{m-1} \sqa^{(m-1)}_{h_{m},\ldots,h_{j+2},h_{j+1}\p h_j,h_{j-1},\ldots,h_1} \quad (m\geq 2).
\eeq
Finally, the conformal differential operators $\t\sqa_h^{(m)}$ are given, for $m\geq 2$, in terms of intermediate operators $\t\sqa_{h_m,\ldots,h_1;y}^{(m)}$ that are multilinear in $h_m,\ldots,h_1$ by
\beq\label{tsqa1}
	\t\sqa^{(m)}_h = \t\sqa^{(m)}_{h,h,\ldots,h;\p h/h}
\eeq
and these intermediate operators satisfy the recursion relations
\beq\label{tsqa2}
	\t\sqa^{(m)}_{h_m\ldots,h_1;y}
	= -\nabla_{h_m}\t\sqa^{(m-1)}_{h_{m-1},\ldots,h_1;y}
	- \sum_{j=1}^{m-1}
	\t\sqa^{(m-1)}_{h_{m},\ldots,h_{j+2},yh_{j+1}h_j,h_{j-1},\ldots,h_1;y} \quad
	(m\geq 2).
\eeq

\end{lemma}
\noindent\proof
The existence of the asymptotic expansion is equivalent to the existence of all derivatives
\[
	\lt.\lt(\frc{d}{d\ep}\rt)^m f(g_\ep)\rt|_{\ep=0},\quad m\geq0.
\]
The existence of these derivatives can be established using the definition of smoothness \cite[Def 2.19]{Dhigher}. We first note that for $\ep\in N$, we have
\[
	\frc{d}{d\ep} g_\ep = h_\ep\circ g_\ep
\]
where $h_\ep = h\circ g_\ep^{-1}$ is holomorphic on $g_\ep(K)$ and $h_{\ep'}\to h_\ep$ as $\ep'\to\ep$ compactly on a neighborhood of $g_\ep(K)$. Hence by differentiability we have
\[
	\frc{d}{d\ep} f(g_\ep) = \nabla_{h_\ep} f(g_\ep)
\]
for all $\ep\in N$. The definition of smoothness implies \cite[Eq 2.8]{Dhigher}, which says that we may differentiate with respect to $\ep$ by differentiating in turn with respect to the linear argument $h_\ep$ of $\nabla_{h_\ep} f(g_\ep)$, and with respect to the argument $g_\ep$ of the derivative itself. By inspection, we find that $h_\ep$ is infinitely-many times differentiable with respect to $\ep$ compactly on a neighborhood of $g_\ep(K)$, and all derivatives can be evaluated using recursively
\[
	\frc{d}{d\ep} h_\ep = -h_\ep \,\p h_\ep.
\]
Hence
\[
	\lt(\frc{d}{d\ep}\rt)^2 f(g_\ep) = -\nabla_{h_\ep \p h_\ep}f(g_\ep) + \nabla_{h_\ep,h_\ep}f(g_\ep),
\]
and continuing this process we find that all derivatives exist.

Let $\sqa^{(m-1)}_h$ be the differential operator obtained by this process after $m-1$ differentiation with respect to $\ep$, for some $m\geq 2$. We see that we may construct $\sqa^{(m)}_h$ from it by adding the term $\nabla_h \sqa^{(m-1)}_h$, and the terms obtained from $\sqa^{(m-1)}_h$ by replacing every factor of $h$ by $-h\p h$. This can be expressed by defining recursively, for $m\geq 2$, intermediate differential operators $\sqa^{(m)}_{h_m,\ldots,h_1}$ via (\ref{sqa2}). Then we simply have (\ref{sqa1}).

The expansion of $f(g_\ep^{-1})$ is obtained in a similar fashion. We first observe that
\[
	\frc{d}{d\ep} g_\ep^{-1} = -\t h_\ep \circ g_\ep^{-1}
\]
where $\t h_\ep = h/\p g_\ep$ is holomorphic on $K$ and $\t h_{\ep'}\to\t h_\ep$ as $\ep'\to\ep$ compactly on a neighborhood of $K$. Further, $\t h_\ep$ is infinitely-many times differentiable with
\[
	\frc{d}{d\ep} \t h_\ep = -\frc{\p h}{h} \t h_\ep^2.
\]
Hence,
\[
	\frc{d}{d\ep} f(g_\ep^{-1}) = -\nabla_{\t h_\ep} f(g_\ep^{-1})
\]
and all other derivatives may be evaluated by differentiating with respect to the linear argument of $\nabla$ and the argument of the derivative itself, as above. This leads to (\ref{tsqa2}) with (\ref{tsqa1}), where the extra index $y$ takes care of the $\ep$-independent factor $\p h/h$.
\eproof

The recursion relations (\ref{sqa2}) and (\ref{tsqa2}) allow us to evaluate all differential operators $\sqa_h^{(m)}$, $\t\sqa_h^{(m)}$ in terms of multiple conformal derivatives $\nabla_{h,h',\ldots} := \nabla_h \nabla_{h'}\cdots$. For instance, we immediately obtain
\beq
	\sqa_h^{(2)} = \nabla_{h,h} - \nabla_{h\p h},\quad
	\t\sqa_h^{(2)} = \nabla_{h,h} + \nabla_{h\p h}
\eeq
as well as
\beqa
	\sqa_h^{(3)} &=&
	\nabla_{h,h,h} - 2 \nabla_{h,h\p h}-\nabla_{h\p h,h}
	+ 2 \nabla_{h (\p h)^2} + \nabla_{h^2\p^2 h} \n
	\t\sqa_h^{(3)} &=& -\nabla_{h,h,h} - 2 \nabla_{h,h\p h}-
	\nabla_{h\p h,h}
	- 2 \nabla_{h (\p h)^2}\ .
\eeqa

We note that the ``symmetrized'' versions, using $[\nabla_h,\nabla_{h'}] = \nabla_{h\p h'-h'\p h}$, are
\beqa
	\sqa_h^{(3)} &=& \nabla_{h,h,h} - \frc32 \lt(\nabla_{h,h\p h}
	+\nabla_{h\p h,h}\rt) + \frc12 \nabla_{h^2\p^2 h} + 2 \nabla_{h(\p h)^2} \n
	\t\sqa_h^{(3)} &=& -\nabla_{h,h,h} - \frc32 \lt(\nabla_{h,h\p h}
	+\nabla_{h\p h,h}\rt) - \frc12 \nabla_{h^2\p^2 h} - 2 \nabla_{h(\p h)^2}\ .
\eeqa
From this one can deduce a simple algorithm that relates symmetrized versions of $\sqa_h^{(m)}$ and $\t\sqa_h^{(m)}$: one simply has to invert the sign of every single-differential operator $\nabla$. We will show this in another publication.

Naturally, expanding $f$ about $\id$ is not essential; we may also expand it about another appropriate conformal map $g$, considering $f(g_\ep\circ g)$ and $f(g_\ep^{-1}\circ g)$. More generally, we may consider functions $f$ on a subset ${\cal N}$ of objects on which there is an action of conformal maps near the identity in the $C$-topology: $\Sigma\mapsto g\cdot\Sigma$ with $(g\circ g')\cdot\Sigma = g\cdot g'\cdot\Sigma$. For instance, ${\cal N}$ may be the set of all compact subsets of $C$, with an action $g\cdot\Sigma = g(\Sigma)$. We have in this context
\beq\label{expgeqSigma}
	f(g_\ep\cdot\Sigma) \sim \sum_{m\geq 0} \frc{\ep^m}{m!} \,\sqa_h^{(m)}f(\Sigma),\quad
	f(g_\ep^{-1}\cdot\Sigma) \sim \sum_{m\geq 0} \frc{\ep^m}{m!} \,\t\sqa_h^{(m)}f(\Sigma)
\eeq
where all derivatives are with respect to the action $\Sigma\mapsto g\cdot\Sigma$ (for instance, $\nabla_h f(\Sigma) = \lt.\frc{d}{dt} f((\id+th)\cdot\Sigma)\rt|_{t=0}$). In particular, from $f(g_\ep\cdot g_\ep^{-1}\cdot\Sigma) = f(\Sigma)$, we have the differential-operator equation
\beq
	\sum_{m=0}^\infty \sum_{m'=0}^\infty \frc{\ep^{m+m'}}{m!\,m'!}
	\sqa_h^{(m)} \t\sqa_h^{(m')} =1,
\eeq
which can be checked explicitly order by oder in $\ep$.

The operators $\t\sqa_h^{(m)}$, the only ones that we will actually use, have a particularly simple form, and we can describe them slightly more explicitly.
\begin{lemma}
We have
\beq\label{tsqa0}
	\t\sqa_h^{(m)} = (-1)^m \sum_\lambda C_\lambda
	\nabla_{h (\p h)^{\lambda_j-1}}\cdots
	\nabla_{h (\p h)^{\lambda_1-1}}
\eeq
where the sum is over all ordered partitions $\lambda=(\lambda_1,\ldots,\lambda_j)$ of $m$ into $j$ parts, for all $j=1,2,3,\ldots,m$ (that is: $\lambda_i\geq 1$ for all $i=1,2,\ldots,j$ and $\sum_{i=1}^j \lambda_i = m$). The coefficients $C_\lambda$ satisfy the following recursion relations
\beq\label{tsqaC}
	C_{(\lambda_1,\ldots,\lambda_j)} =
	\delta_{\lambda_j,1} C_{(\lambda_1,\ldots,\lambda_{j-1})}
	+ \sum_{i=1}^j (\lambda_i-1)
	C_{(\lambda_1,\ldots,\lambda_i-1,\ldots,\lambda_j)}
\eeq
for all $j\geq 1$, with initial condition $C_{(1)} = 1$.
\end{lemma}
\proof From (\ref{tsqa2}), it is clear that all terms that occur in the intermediate multilinear operators are products of single derivatives of the form $\nabla_{y^n h_{j+n}\ldots h_j}$, which become under (\ref{tsqa1}) of the form $\nabla_{h (\p h)^n}$. Further, it is also clear from (\ref{tsqa2}) that all combinations appear, as long as there are exactly $m$ factors of $h$ or its derivatives; this lead to the form (\ref{tsqa0}). Finally, for every term corresponding to the partition $\lambda$, there are two types of sources to the recursion relation for the coefficient $C_\lambda$: that coming from the first term in (\ref{tsqa2}), giving immediately the first term in (\ref{tsqaC}); and that coming from the second term in (\ref{tsqa2}). Concerning the latter one, we see that every term with a factor $\nabla_{h (\p h)^n}$ leads, at the next order, to exactly $n+1$ times the same term with that factor replaced by $\nabla_{h(\p h)^{n+1}}$, and this for every such factor, which gives rise to the second term in (\ref{tsqaC}).
\eproof

This shows (\ref{Clambda}). In order to show (\ref{Clambda2}), we count as follows. Start with $C_{(\lambda_1,\ldots,\lambda_j)}$ and consider using the second term in (\ref{tsqaC}) recursively in order to reach $C_{(\ell_1,\ldots,\ell_{j-1},1)}$. This produces a factor $p = \prod_{i=1}^j \frc{(\lambda_i-1)!}{(\ell_i-1)!}$, and is done in exactly $\sum_i(\lambda_i-\ell_i)$ steps, where $\ell_j:=1$. Amongst these steps, we need to choose, for every $i=1,\ldots,j$, exactly $\lambda_i-\ell_i$ of them where it is the $i^{\rm th}$ factor that is decreased. There are $q= \frc{\lt(\sum_i(\lambda_i-\ell_i)\rt)!}{\prod_{i=1}^j (\lambda_i-\ell_i)!}$ ways of doing this. Hence, we obtain $pq C_{(\ell_1,\ldots,\ell_{j-1},1)}$. From there, the two terms in (\ref{tsqaC}) may be used. The use of the second term is taken into account by the above counting for reaching $C_{(\ell_1,\ldots,\ell_{i}-1,\ldots,\ell_{j-1},1)}$. Hence, we only use the first term, and sum over all $(\ell_1,\ldots,\ell_{j-1})$. The first term leads to $pq C_{(\ell_1,\ldots,\ell_{j-1})}$. This shows (\ref{Clambda2}). The initial condition $C_{(n)} = (n-1)!$ is obvious from (\ref{tsqaC}).

Finally, in order to show (\ref{Clambda3}), we proceed as follows. It is obvious from (\ref{tsqaC}) that $C_{(1,\ldots,1)} = 1$, and hence, from recursive use (\ref{Clambda2}), that $C_{(1,\ldots,1,\lambda_1,\ldots,\lambda_j)} = C_{(\lambda_1,\ldots,\lambda_j)}$. The recursion relation (\ref{Clambda2}) gives rise to
\beq\label{recn}
	C_{(n,1_k)} = \sum_{\ell=1}^n \frc{(n-1)!}{(\ell-1)!} C_{(\ell,1_{k-1})}
\eeq
where $1_k = 1,1,\ldots,1$ ($k$ times). Using the formula
\[
	\sum_{\ell=1}^n \frc{(\ell+k-1)!}{(\ell-1)!} = \frc{k! (n+k)!}{(n-1)!(k+1)!}
\]
we see that the second equation of (\ref{Clambda3}) indeed solves (\ref{recn}), with the correct initial condition.

In (\ref{gep}), we may consider $\ep$ to be a complex number as well, and we may look for an expansions of $f(g_\ep)$ and of $f(g_\ep^{-1})$ in powers of $\ep$ and of its complex conjugate $\b \ep$. These expansions exist, and we have the following.
\begin{lemma}\label{lemexpholo}
In the context of Lemma \ref{lemexpf}, the functions $f(g_\ep)$ and $f(g_\ep^{-1})$ have asymptotic expansions in nonnegative integer powers of both $\ep$ and $\b\ep$. The coefficients $\sqa^{(m)}[h] f(\id)$, $\t\sqa^{(m)}[h] f(\id)$ and $\underline{\sqa}^{(m)}[\b h] f(\id)$, $\underline{\t\sqa}^{(m)}[\b h] f(\id)$ of pure $\ep$ and $\b\ep$ powers respectively,
\beqa
	f(g_\ep) &\sim& \sum_{m\geq0} \frc{\ep^m}{m!} \sqa^{(m)}[h] f(\id)
	+ \sum_{m\geq0} \frc{\b\ep^m}{m!} \underline\sqa^{(m)}[\b h] f(\id)
	+ \mbox{mixed} \n
	f(g_\ep^{-1})
	&\sim& \sum_{m\geq0} \frc{\ep^m}{m!} \t\sqa^{(m)}[h] f(\id)
	+ \sum_{m\geq0} \frc{\b\ep^m}{m!}
	\underline{\t\sqa}^{(m)}[\b h] f(\id) + \mbox{mixed},\no
\eeqa
are obtained by replacing in $\sqa^{(m)}_h$, $\t\sqa^{(m)}_h$ every single-derivative operators $\nabla_h$ by holomorphic derivative $\Delta[h]$, respectively anti-holomorphic derivative $\b\Delta[\b h]$.
\end{lemma}
\proof This is a simple consequence of Lemma \ref{lemexpf}, the definitions (\ref{Deltah}) of holomorphic and anti-holomorphic derivatives, and the fact that $\Delta[\cdot]$ and $\b\Delta[\cdot]$ are $\C$-linear \cite{Dhigher}. \eproof

Note that $\sqa^{(m)}[\cdot]$, $\t\sqa^{(m)}[\cdot]$, $\underline{\sqa}^{(m)}[\cdot]$ and $\underline{\t\sqa}^{(m)}[\cdot]$ are all $m$-$\C$-linear.

A set of functions $h$ which are of particular interest for deriving our main result are the functions (\ref{h}), or rather,
\beq\label{hh}
	\h h_{k,w} = -h_{-k,w}.
\eeq
It is a simple matter, using $\h h_{k,w} (\p \h h_{k,w})^{n-1} = -(1-k)^{n-1} h_{-kn,w}$ in (\ref{tsqa0}) and using Lemma \ref{lemexpholo}, to find
\beq\label{restsqa}
	\t\sqa^{(m)}[\h h_{k,w}] = \sum_{\lambda} C_\lambda (k-1)^{m-j}
	\Delta[{h_{-k\lambda_j,w}}]\cdots \Delta[{h_{-k\lambda_1,w}}].
\eeq
For the purpose of our construction, this is the most important result of this section.

\subsection{Proofs of the main results}

Our proofs of Theorems \ref{res2} and \ref{res3} follow very closely the steps of the proofs provided in \cite{DTCLE}, but are here expressed in the language of a general conformal restriction system.

The hypotrochoids (\ref{higher}) are the boundaries of domains $\hC\setminus \cl{E_k(w,\ep,\theta,b)}$ that are images under
\beq\label{gk}
	g_{k,w,\ep,\theta}(z) = z + \ep^{k} e^{ki\theta} \h h_{k,w}(z)
	= z + \frc{\ep^k e^{ki\theta}}{(z-w)^{k-1}}
\eeq
of the region $\hC\setminus (b\ep \cl\uD+w)$, the outside of the disk of radius $b\ep$ centered at $w$.

The proof proceeds by induction on the number of insertions of hypotrochoid fields.

Consider the completion $\cl{\h{\frak{X}}}$ of the linear space $\h{\frak{X}}$ (recall Subsection \ref{ssectDefCR}) under the weak-local topology. Since the weak-local convergence condition in $\h{\frak{X}}$ is stronger than that in $\frak{X}$, we have $\cl{\h{\frak{X}}}\subset \cl{\frak{X}}$. In general, $\cl{\h{\frak{X}}}$ is not an algebra, but by definition of weak-local convergence, if $\tY\in \cl{\h{\frak{X}}}$ and $\tX\in\h{\frak{X}}$ have disjoint supports, then $\tX\tY\in\cl{\h{\frak{X}}}$. When seeing $\tX\tY$ as an element in $\cl{\frak{X}}$, the order of the limits is, in general, important. More precisely, assume $\lim_{i\to\infty} \tY_i=\tY$ weakly locally in $\h{\frak{X}}$. Then the product $\tX\tY\in\cl{\frak{X}}$ is defined by taking first, if need be, the limits $N\to\alpha$ defining the factors $\tE(\alpha)$ eventually present in $\tX$ and in $\tY_i$ (so that we get elements in $\h{\frak{X}}$), and then the limit $i\to\infty$ of $\tX\tY_i$. Note that the Restriction property holds for every $\tX\in\cl{\h{\frak{X}}}$, indeed one just has to take the limit on both sides of (\ref{restr}), which exists by weak-local convergence.

Let $\t{\frak{X}}\subset \cl{\h{\frak{X}}}$ be a linear subspace containing $\h{\frak{X}}$ that carries a representation of the groupoid of conformal maps, and such that Conformal invariance, Smoothness and Local covering hold for every $\tX\in\t{\frak{X}}$ (again, here we require only their validity for functionals $\mathbb{E}\big[\cdot\big]_A$ on regions $A$ whose boundary components, if any, are Jordan curves).

In the first step of the induction, we take $\t{\frak{X}} = \h{\frak{X}}$.

Let $C$ be a region with Jordan boundary components $\p_i C$. We admit $C=\hC$, in which case the set $\{\p_i C\}$ is empty. Let $w\in C$, and let us denote by $g_{k,w,\ep,\theta}^{-1}\cdot C$ the domain bounded by $g_{k,w,\ep,\theta}^{-1}(\p C)$ and containing $w$ (for $\ep$ small enough). In particular, $g_{k,w,\ep,\theta}^{-1}\cdot \hC = \hC$. Let $\tY\in\t{\frak{X}}$ be supported in $C\setminus \{w\}$, and let us consider the expectation $Y:=\mathbb{E}\big[\tY]_{C\setminus (b\ep\cl\uD+w)}$ as a function of conformal maps $g$ using $\tY\mapsto g\cdot \tY$ and $C\mapsto g\cdot C$. Here, we take all $g$ in a neighborhood of the identity with respect to the topology of \cite{Dhigher}, with conformal maps on a closed annular set containing $\p C$ and a support of $\tY$. Let us introduce the operator ${\rm T}$ (resp.~${\rm T}^{-1}$) which ``translates'' by the conformal map $g_{k,w,\ep,\theta}$ (resp.~$g_{k,w,\ep,\theta}^{-1}$). Then we have
\beq\label{eqTm1}
	{\rm T}^{-1}\lt(Y\rt) =
	\mathbb{E}\big[g_{k,w,\ep,\theta}^{-1} \cdot \tY\big]_{
	g_{k,w,\ep,\theta}^{-1}\cdot C \setminus (b\ep \cl\uD+w)}
	= \mathbb{E}\big[\tY\big]_{C\setminus \cl{E_k(w,\ep,\theta,b)}}
	= \frc{\mathbb{E}\big[\tE_k(w,\ep,\theta,b)\,\tY\big]_C}{
	\mathbb{E}\big[\tE_k(w,\ep,\theta,b)\big]_C}
\eeq
for all $\ep$ small enough, where in the second equality we used Conformal invariance and in the third, Restriction.

Let the simply connected Jordan region $D\ni w$ be such that $\cl D\subset C$. Recall the notation of Subsection \ref{ssectImplic} introduced above (\ref{relaa}). With $C$ and $D$ transforming as $C\mapsto g\cdot C$ and $D\mapsto g\cdot D$ under conformal maps on $\p C$ and $\p D$, respectively, let $U:=\mathbb{E}\big[\prod_i \tE(\p_i C)\big]_{\hC}$, $V:=\mathbb{E}\big[\prod_i \tE(\p_i C)\big]_{\hC\setminus \cl D}$, $W:=\mathbb{E}\big[\prod_i \tE(\p_i C )\big]_{\hC\setminus (b\ep \cl\uD+w)}$, and $X:=\mathbb{E}\big[\tE_k(w,\ep,\theta,b)\big]_C$ be functions of conformal maps $g$ near the identity. We take $W=V=U=1$ if $C=\hC$. Then (\ref{relaa}) and the fact that $\mathbb{E}\big[\tE_k(w,\ep,\theta,b)\big]_\hC = 1$ imply
\beq\label{UX}
	UX = \mathbb{E}\lt[\prod_i \tE(\p_i C)\rt]_{\hC\setminus \cl{E_k(w,\ep,\theta,b)}}.
\eeq
We may now evaluate the factor $F(g,A)$ in (\ref{transfoE}) in two different ways:
\beq
	f(g_{k,w,\ep,\theta},\hC\setminus \cl C) = \frc{{\rm T}(V)}{V} = \frc{{\rm T}(UX)}{W}
\eeq
where in the last equation we used (\ref{UX}). This implies that
\beq
	X = \frc{V}U {\rm T}^{-1} \lt(\frc{W}V\rt).
\eeq
With (\ref{eqTm1}), we find
\beq
	\mathbb{E}\big[\tE_k(w,\ep,\theta,b)\,\tY\big]_C
	= X {\rm T}^{-1}\lt( Y\rt)
	= \frc{V}U {\rm T}^{-1} \lt(\frc{WY}V
	\rt).
\eeq

We see that the relative partition function (\ref{Zgen}) is simply $Z(\p C,\p D) = U/V$. By Smoothness, Lemma \ref{lemexpholo} and Eq.~(\ref{gk}), we then obtain
\beqa
	\lefteqn{\mathbb{E}\big[\tE_k(w,\ep,\theta,b)\,\tY\big]_C
	}&& \n &=& Z(\p C,\p D)^{-1} \sum_{m\geq 0}\lt( \frc{\ep^{km} e^{kmi\theta}}{m!}
	\t\sqa^{(m)}[\h h_{k,w}] + 
	\frc{\ep^{km} e^{-kmi\theta}}{m!}
	\underline{\t\sqa}^{(m)}[\h h_{k,w}] \rt)\lt(\frc{WY}V
	\rt) \n && +
	\sum_{m\in\Z} e^{kmi\theta} o(\ep^{km}). \label{expsa}
\eeqa
(Anti-)holomorphicity in $w$ of the intervening multiple (anti-)holomorphic conformal derivatives \cite{Dcalc,Dhigher} then shows Theorem \ref{res2}.

Note that Local covering implies $\lim_{\ep\to0} W = U$ and $\lim_{\ep\to0}Y =\mathbb{E}\big[\tY]_C$, and further that conformal differential operators commute with the limit operation. Hence, we may replace $W/V$ by $Z(\p C,\p D)$ and $Y$ by $\mathbb{E}\big[\tY]_C$ in (\ref{expsa}), so that we have
\beqa
	\lefteqn{\mathbb{E}\big[\tE_k(w,\ep,\theta,b)\,\tY\big]_C
	}&& \n &=& Z(\p C,\p D)^{-1} \sum_{m\geq 0}\lt( \frc{\ep^{km} e^{kmi\theta}}{m!}
	\t\sqa^{(m)}[\h h_{k,w}] + 
	\frc{\ep^{km} e^{-kmi\theta}}{m!}
	\underline{\t\sqa}^{(m)}[\h h_{k,w}] \rt)\lt(Z(\p C,\p D)
	\mathbb{E}\big[\tY]_{C}
	\rt) \n && +
	\sum_{m\in\Z} e^{kmi\theta} o(\ep^{km}). \label{expsafin}
\eeqa

Along with (\ref{restsqa}), Equation (\ref{expsafin}) shows (\ref{eval}) (weakly locally, and more generally for $\tX\in\t{\frak{X}}$). Equation (\ref{eval}) gives rise to transformation properties for the hypotrochoid fields which agree with the transformation properties found from CFT, thanks to \cite[Prop 3.9, Thm 3.3]{Dhigher}\footnote{Note in particular that thanks to factorization of conformal maps on finitely connected regions \cite{fact1,fact2,Dfact}, the results of \cite{Dhigher} can be generalized to transformation properties for conformal maps on finitely connected regions, as in the proof of Theorem \ref{theoZgen}.} and Theorem \ref{theoZgen}. Let us now consider $\t{\frak{X}}'$, where all elements of the form $\tY \tT_{k,m}(w'),\;w'\in\C,\;\tY\in\t{\frak{X}}$ supported in $\hC\setminus \{w'\}$ (the limit defining $\tT_{k,m}(w')$ taken last) have been adjoined to $\t{\frak{X}}$. The transformation properties give $\t{\frak{X}}'$ the structure of a representation of the groupoid of conformal maps, such that Conformal invariance holds. By inspection of the explicit form (\ref{eval}), and of the explicit transformation property of \cite{Dhigher}, we further conclude that both Smoothness and Local covering hold. Hence, we may repeat the induction process with $\t{\frak{X}}'$ in place of $\t{\frak{X}}$.

This induction process shows that the general expression for multiple insertions of hypotrochoid fields yields the multiple-conformal-derivative expression obtained by recursive use of (\ref{eval}), where the order of the differential operators is tied with the order in which the limits defining the fields hypotrochoid fields are taken. Thanks to \cite[Thm 4.2]{Dhigher}, this gives rise to (\ref{voaeval}). Then, the commutativity property of vertex operator algebras \cite{LL04} implies that the limits can be taken in any order giving the same results, hence that the hypotrochoid fields form a consistent set with respect to $\h{\frak{X}}$. This completes the proof.

\sect{Conclusion} \label{sectConclusion}

We have studied certain renormalized random variables in conformal restriction systems, CLE being expected to give rise to an example of such a system, and identified them with descendants of the identity field in CFT. Our proofs involved two main steps. We first analyzed the expansion in $\ep$ of $f(\id+\ep h)$ for general smooth functions $f$, in positive integer powers of $\ep$, and established a relation between the coefficients and particular multiple conformal derivatives. This is useful, because our recent work \cite{Dhigher} provides the vertex operator algebraic structure of multiple conformal derivatives, and \cite{Dcalc} relates conformal derivatives to the conformal Ward identities of CFT. We then used this in conjunction with the axioms of conformal restriction systems in order to derive the main results.

It would be most interesting to extend this work to other curves than those of hypotrochoid type. These curves lead to simple fixed-spin holomorphic fields, but it would be interesting to have the general description for arbitrary curves. On the other hand, it would of course be interesting to have the full Virasoro vertex operator algebra in terms of similar geometric objects in conformal restriction systems. In particular, an interesting question is about the geometric meaning of the infinitely-many higher-spin conserved densities of CFT.

{\bf Acknowledgments.}
I thank the Galileo Galilei Institute for Theoretical Physics for the hospitality and the INFN for partial support during the completion of this work, as well as SISSA, Trieste for hospitality. I also thank J. Cardy for comments on the manuscript. This work was supported by EPSRC under grant EP/H051619/1 ``From conformal loop ensembles to conformal field theory" (First Grant scheme).


\begin{thebibliography}{99}

\bibitem{BBSLE} Bauer, M. and Bernard, D.: 2D growth processes: SLE and Loewner chains. {\tt arXiv:math-ph/0602049}, Phys. Rep. 432, 115 (2006).

\bibitem{BPZ} Belavin, A. A., Polyakov, A. M., Zamolodchikov, A. B.: Infinite conformal symmetry in two-dimensional quantum field theory. Nucl. Phys. B 241, 333 (1984)

\bibitem{Cardy05} Cardy, J.: SLE for theoretical physicists. {\tt arXiv:cond-mat/0503313} Ann. of Phys. 318(1), 81 (2005)

\bibitem{DFMS97} Di Francesco, P., Mathieu, P., Senechal, D.: {\em Conformal Field Theory}, Berlin, Springer, 1997

\bibitem{Dcalc} Doyon, B: Calculus on manifolds of conformal maps and CFT. {\tt arXiv:1004.0138}, J. Phys. A 45 (2012) 315202

\bibitem{Dfact} Doyon, B: Factorization of conformal maps on finitely connected domains, {\tt arXiv:1107.0582}.

\bibitem{Dhigher} Doyon, B: Higher conformal variations and the Virasoro vertex operator algebra. {\tt arXiv:1110.1507}

\bibitem{DTCLE} Doyon, B: Conformal loop ensembles and the stress-energy tensor. {\tt arXiv:1209.1560}, to appear in Lett. Math. Phys.

\bibitem{DRC06} Doyon, B., Riva, V., Cardy, J.: Identification of the stress-energy tensor through conformal restriction in SLE and related processes. {\tt arXiv:math-ph/0511054}, Comm. Math. Phys. 268, 687 (2006)

\bibitem{FW03} Friedrich, R., Werner, W.: Conformal restriction, highest-weight representations and SLE. {\tt arXiv:math-ph/0301018}, Comm. Math. Phys. 243 (1), 105 (2003)

\bibitem{Gins} Ginsparg, P.: Applied conformal field theory, in: Les Houches, session XLIX (1988), {\em Champs, cordes et ph\'enom\`enes critiques / Fields, strings and critical phenomena}, Eds. E. Br\'ezin and J. Zinn-Justin, Elsevier, New York (1989)

\bibitem{GSLE} I. A. Gruzberg: Stochastic geometry of critical curves, Schramm-Loewner evolutions, and conformal field theory. {\tt  arXiv:math-ph/0607046}, J. Phys. A 39, 12601 (2006).

\bibitem{fact1} H\"ubner, O: Die Faktorisierung konformer Abbildungen und Anwendungen, Math. Zeitschr. 92 (1966) 95.

\bibitem{fact2} K\"uhnau, R: Einige elementare Bemerkungen zur Theorie der konformen und quasikonformen Abbildungen, Math. Nach. 45 (1970) 307.

\bibitem{KNSLE} Kager, W. and Nienhuis, B.: A Guide to Stochastic Loewner Evolution and its Applications. {\tt arXiv:math-ph/0312056}, J. Stat. Phys. 115, 1149 (2004).

\bibitem{LSLE} Lawler, G. F.: Conformally invariant processes in the plane. Mathematical Surveys and Monographs, 114. American Mathematical Society, Providence, RI, 2005.

\bibitem{LL04} Lepowsky, J., Li, H.: {\em Introduction to Vertex Operator Algebras and Their Representations}. Progress in Mathematics, Vol. 227, Boston, Birkh\"auser, 2004

\bibitem{N82} Nienhuis, B.: Exact critical point and critical exponents of $O(n)$ models in two dimensions. Phys. Rev. Lett 49, 1062 (1982)

\bibitem{RBGWSLE} Rushkin, I., Bettelheim, E., Gruzberg, I. A. and Wiegmann, P.: Critical curves in conformally invariant statistical systems. {\tt arXiv:cond-mat/0610550}, J. Phys. A 40, 2165 (2007).

\bibitem{S00} Schramm, O.: Scaling limits of loop-erased random walks and uniform spanning trees. {\tt arXiv:math.PR/9904022}, Israel J. Math. 118, 221 (2000)

\bibitem{Sh06} Sheffield, S.: Exploration trees and conformal loop ensembles. {\tt arXiv:math.PR/0609167}, Duke Math. J. 147, 79 (2009)

\bibitem{ShW07} Sheffield, S., Werner, W.: Conformal loop ensembles: The Markovian characterization and the loop-soup construction. {\tt arXiv:1006.2374}, to appear in Ann. Math.

\bibitem{Smi1} Smirnov, S.: Critical percolation on the plane: conformal invariance, Cardy's formula, scaling limits. C. R. Acad. Sci. Paris S\'er. I Math., 333(3), 239 (2001);

\bibitem{Smi2} Smirnov, S.: Towards conformal invariance of 2D lattice models. {\tt arXiv:0708.0032}, {\em Proceedings of the International Congress of Mathematicians, Madrid 2006}, Vol. II., 1421-1451, Zurich, Eur. Math. Soc. (2006)

\bibitem{Smi3} Smirnov, S.: Conformal invariance in random cluster models. I. Holomorphic fermions in the Ising model. {\tt arxiv:0708.0039}, Ann. Math. 172, 1435 (2010)

\bibitem{WSLE} Werner, W.: Random planar curves and Schramm-Loewner evolutions. {\tt arXiv:math.PR/0303354}, Lecture Notes in Mathematics 1840, Springer- Verlag (Berlin, 2004) 






\end{thebibliography}
\end{document}